\documentclass[12pt]{amsart}

\usepackage{color}
\usepackage{graphicx, tikz}
\usepackage{xcolor}
\usepackage{cancel}
\usepackage{amssymb,amsfonts,amsthm,amsmath,calligra, enumerate, stmaryrd}
\usepackage{extarrows}
\usepackage[utf8]{inputenc}
\usepackage{accents}
\usepackage{slashed}
\usepackage{yfonts}
\usepackage{mathrsfs,pifont}
\usepackage{mathtools}
\usepackage{todonotes}

\usepackage[bookmarksnumbered=true, bookmarksopen=true]{hyperref}
\usepackage[margin=1in]{geometry}
\usepackage[all]{xy}

\usepackage{aliascnt}
\usepackage[nameinlink,noabbrev]{cleveref}

\usepackage{preambles_Saebyeok}

\usepackage{float,circuitikz,ytableau,youngtab,todonotes}
\usetikzlibrary{patterns,shapes.misc}

\theoremstyle{definition}

\newtheorem{Theorem}{Theorem}[section]

\newaliascnt{Proposition}{Theorem}
\newtheorem{Proposition}[Proposition]{Proposition}
\aliascntresetthe{Proposition}

\newaliascnt{Lemma}{Theorem}
\newtheorem{Lemma}[Lemma]{Lemma}
\aliascntresetthe{Lemma}

\newaliascnt{Corollary}{Theorem}

\aliascntresetthe{Corollary}

\newaliascnt{Definition}{Theorem}
\newtheorem{Definition}[Definition]{Definition}
\aliascntresetthe{Definition}

\newaliascnt{Example}{Theorem}

\aliascntresetthe{Example}

\newaliascnt{Remark}{Theorem}
\newtheorem{Remark}[Remark]{Remark}
\aliascntresetthe{Remark}

\crefname{Theorem}{Theorem}{Theorems}
\crefname{Proposition}{Proposition}{Propositions}
\crefname{Lemma}{Lemma}{Lemmas}
\crefname{Corollary}{Corollary}{Corollaries}
\crefname{Definition}{Definition}{Definitions}
\crefname{Remark}{Remark}{Remarks}
\crefname{section}{Section}{Sections}
\crefname{subsection}{Section}{Sections}
\crefname{appendix}{Appendix}{Appendices}
\crefname{figure}{Figure}{Figures}
\makeatletter
\AddToHook{cmd/appendix/before}{\def\cref@section@alias{appendix}}
\AddToHook{cmd/appendix/before}{\def\cref@subsection@alias{appendix}}
\makeatother

\numberwithin{equation}{section}

\def\ben{\begin{eqnarray*}}
\def\een{\end{eqnarray*}}

\newcommand{\bQ}{\mathbb{Q}}

\newcommand{\bT}{\mathbb{T}}

\newcommand{\bt}{\mathbf{t}}

\makeatletter
\newcommand{\ostar}{\mathbin{\mathpalette\make@circled\star}}
\newcommand{\make@circled}[2]{%
  \ooalign{$\m@th#1\smallbigcirc{#1}$\cr\hidewidth$\m@th#1#2$\hidewidth\cr}%
}
\newcommand{\smallbigcirc}[1]{%
  \vcenter{\hbox{\scalebox{0.77778}{$\m@th#1\bigcirc$}}}%
}
\makeatother

\hypersetup{colorlinks,breaklinks,
            linkcolor = purple,
            citecolor = teal,
            urlcolor = purple,
            }

\def\qe{\mathfrak{q}}
\def\kq{\mathfrak{q}}

\def\by{\mathbf{y}}

\def\bx{\mathbf{x}}
\def\ba{\mathbf{a}}
\def\bm{\mathbf{m}}

\def\bt{\mathbf{t}}

\def\bT{\mathbf{T}}

\def\bQ{\mathbf{Q}}

\def\BC{\mathbb{C}}
\def\BR{\mathbb{R}}

\def\BZ{\mathbb{Z}}

\def\CalN{\mathcal{N}}

\def\CalV{\mathcal{V}}

\def\CalZ{\mathcal{Z}}
\def\CalH{\mathcal{H}}

\def\CalS{\mathcal{S}}
\def\CalC{\mathcal{C}}
\def\CalQ{\mathcal{Q}}
\def\CalT{\mathcal{T}}
\def\CalF{\mathcal{F}}
\def\qe{\mathfrak{q}}

\def\Tr{\operatorname{Tr}}
\def\Ad{\operatorname{Ad}}

\def\ve{{\varepsilon}}

\def\ta{\mathtt{a}}

\def\EZ{\EuScript{Z}}

\def\EH{\EuScript{H}}
\def\EV{\EuScript{V}}

 \def\p{\partial}
 
 \def\a{\alpha}
 \def\b{\beta}
 \def\g{\gamma}
 \def\d{\delta}

 \def\l{\lambda}
 \def\m{\mu}

 \def\s{\sigma}
 \def\t{\tau}

 \def\G{\Gamma}
 \def\D{\Delta}
 \def\T{\Theta}

 \def\S{\Sigma}
 \def\L{\Lambda}
 \def\O{\Omega}
 \def\o{\omega }

  \def\bve{\boldsymbol\ve}
  \def\bom{\boldsymbol\omega}

\def\beq{\begin{equation}}
\def\eeq{\end{equation}}
\title{$Q$-operators, cluster mutations, and flavored Toda chain}

\author{Saebyeok Jeong}
\address{Center for Geometry and Physics, Institute for Basic Science (IBS), Pohang 37673, South Korea}
\email{saebyeok.jeong@gmail.com}

\author{Taro Kimura}
\address{Université Bourgogne Europe, CNRS, IMB UMR 5584, 21000 Dijon, France}
\address{Institut Universitaire de France (IUF), France}
\email{taro.kimura@ube.fr}

\author{Norton Lee}
\address{Center for Geometry and Physics, Institute for Basic Science (IBS), Pohang 37673, South Korea}
\email{norton.lee@ibs.re.kr}

\begin{document}

\begin{abstract}
We construct Baxter $Q$-operators for a family of quantum cluster integrable systems, that we call the flavored Toda chains, associated with quivers $\CalQ_N[F]$, where $F\subseteq\{2,\dots,N\}$ specifies data of $\vert F \vert$ additional nodes compared to $\CalQ_N[\varnothing]$ for the ordinary $N$-particle open $q$-deformed Toda chain. 
A distinguished sequence of quantum cluster mutations yields an ordered product of Faddeev's quantum dilogarithms.
We prove that these operators form a commuting family and that their shifted quotients are transfer matrices whose coefficients are commuting Hamiltonians.
We identify these transfer matrices with matrix elements of products of explicit local Lax matrices of rank two.
Local cluster transformations move flavor insertions along the chain while preserving their ordered flavor parameters and the transfer matrix together with the canonical transformation.
We also derive integral kernels for the $Q$-operators and evaluate an associated Fourier transform as a factorized scalar $Q$-function.
Finally, we relate the quivers to the BPS quivers of five-dimensional $\CalN=1$ gauge theories in the decoupling limit and identify the normalized scalar $Q$-function with the perturbative contribution of a canonical codimension-two defect wrapping a squashed three-sphere.
\end{abstract}

\maketitle

\setcounter{tocdepth}{1}
\tableofcontents

\setlength{\parskip}{1ex}
% Allow difficult paragraphs to wrap within the template margins.
\setlength{\emergencystretch}{2em}

\allowdisplaybreaks

%\vspace{1cm}

\section{Introduction and summary}\label{sec:introduction}

Quantum cluster mutations provide explicit transformations of
noncommutative algebras associated with quivers. Their realization
by quantum dilogarithms relates the combinatorics of mutations to
the construction of operators in quantum integrable systems.
This relation appears in the theory of cluster integrable systems
associated with dimer models~\cite{Goncharov:2011hp}, and in the
cluster realization of the Baxter operator for the open $q$-deformed Toda chain developed by Schrader and
Shapiro~\cite{schrader2018b}.
In the latter construction, a distinguished mutation sequence
produces the Baxter operator, while identities of the quantum
dilogarithm control its algebraic and analytic properties.

The purpose of this paper is to extend this construction to a new family of open chains with additional parameters, which we call the \emph{q-deformed flavored Toda chain.} 
We construct their $Q$-operators from quantum cluster mutations, obtain the commuting Hamiltonians and local Lax matrices, and describe cluster transformations relating different positions of the flavors.
We also evaluate a Fourier transform associated with the
$Q$-operator and compare its factorization with the perturbative
contribution of a canonical surface defect in five-dimensional
supersymmetric gauge theory.

\medskip

Set $N \in \mathbb{Z}_{>1}$ and a subset $F\subseteq\{2,\dots,N\}$, and introduce flavor parameters $\{m_k\}_{k\in F} \in \mathbb{R}^{|F|}$. 
We associate with these data a quiver $\CalQ_N[F]$. See \cref{sec:Baxter Q-operator} for details.
The fully \emph{flavored} quiver is
$\CalQ_N=\CalQ_N[\{2,\dots,N\}]$, and the remaining quivers are obtained from the full one by merging processes.

We write $\mathbf{x} = (x_1,\ldots,x_N) \in \mathbb{R}^N$.
For $f \in L^2 (\BR^N)$, we consider the canonical operators
\[
\hat{x}_k f(\bx)=x_k f(\bx),
\qquad
\hat{p}_k f(\bx)=\frac{1}{2\pi i}\frac{\partial f(\bx)}{\partial x_k},
\qquad
[\hat{p}_j,\hat{x}_k]=\frac{\delta_{jk}}{2\pi i}.
\]
By abuse of notation, we often omit the hat symbol when no confusion can arise.
Let $b \in \mathbb{R}_{>0}$ and let $\varphi$ denote Faddeev's quantum dilogarithm function.
See \cref{sec:functions} for details.
For $k=2,\dots,N$, we consider the following operators acting on the functional space $\mathcal{F}_N \subset L^2(\mathbb{R}^N)$ (see \cref{sec:int_op} for details), 
\begin{align*}
T_k(u)
&=
\varphi(x_k-x_{k-1}+u-m_k)^{\mathbf{1}_F(k)}
\varphi(p_k+x_k-x_{k-1}+u)
\varphi(p_k+u),
\end{align*}
where $\mathbf{1}_F$ is the indicator function of $F$. 

Our first result identifies the operator associated with the
distinguished mutation sequence and establishes its commutativity.

\begin{Theorem}[\cref{prop:Baxter-Q,lem:QQ}] 
\label{thm:intro-Q}
The $Q$-operator for $\CalQ_N[F]$ is defined by the \emph{follow the arrow} mutation sequence (see \cref{def:Baxter-Q} and \cref{def:follow the arrow}),
\begin{align}\label{eq:intro-Q}
\bQ_N(u)
&=
\varphi(p_1+u)T_2(u)\cdots T_N(u).
\end{align}
Then, we have 
\begin{align*}
\bQ_N(u)\bQ_N(v)
&=
\bQ_N(v)\bQ_N(u),
\qquad u,v\in\BR.
\end{align*}
\end{Theorem}

From the analogy with the integrable system, we call the argument of $\mathbf{Q}_N$ the spectral parameter.
Thus the mutation sequence determines a commuting family of operators depending on a continuous spectral parameter.
The factorization~\eqref{eq:intro-Q} also displays the effect of
each flavor insertion: it contributes one additional quantum
dilogarithm at the corresponding site.
When $F=\varnothing$, this construction reduces to the Baxter
operator of the $q$-deformed open Toda chain
in~\cite{schrader2018b}. 

To formulate the Hamiltonian statement algebraically, set
\begin{align*}
&q=e^{\pi i b^2} \in \mathbb{C}^\times, \\
&P_k=e^{2\pi b p_k}, 
\qquad
X_k=e^{2\pi b x_k}, \qquad k = 1, \ldots, N, \\
&M_l = e^{2\pi b m_l} ,\qquad l\in F.
\end{align*}
Let us denote the quantum torus algebra for the quiver $\CalQ_N[F]$ by $\CalT_N [F]$ (see \cref{sec:qcm}).
Through \emph{fast forward algorithm} \cite{Franco:2005rj} we identify its generators $P_k^{\pm1}$, $X_k^{\pm1}$, and $M_l^{\pm 1}$, subject to the relations 
\begin{align*}
&P_jP_k=P_kP_j,
\qquad
X_jX_k=X_kX_j,
\qquad
P_jX_k=q^{-2\delta_{jk}}X_kP_j,
\qquad 
j, k = 1, \ldots, N, \\
&M_l \text{ commutes with all generators.}
\end{align*}
Mixed exponentials are Weyl ordered; in particular,
\[
e^{2\pi b(p_k+x_k)}=qP_kX_k.
\]

For a subset $F \subseteq [2,\ldots,N]$ and $1\leq k\leq N$, let 
\[
\bQ_k(u) =\varphi(p_1+u)T_2(u)\cdots T_k(u),
\]
with the product empty when $k=1$.
The shifted quotients of these operators will be understood
algebraically, by applying the functional equations of the
quantum dilogarithm.
This convention separates the following quantum-torus relations from questions about domains of operators. 

\begin{Theorem}[\cref{thm:Lax_matrices,thm:integrability}]
\label{thm:intro-transfer}
Define the Lax matrices,
\begin{align*}
L_k(u) = \begin{pmatrix}
        1 + e^{2\pi b(p_k+u)} & e^{2\pi b(p_k+x_k+u)} + \mathbf{1}_F(k) e^{2\pi b(x_k+u-m_{k})} \\
        e^{-2\pi b x_k} & \mathbf{1}_F(k) e^{2\pi b(u-m_{k})}
\end{pmatrix} \in \text{End}(\BC^2)\otimes \CalT_N[F] .
\end{align*}
For every $1\leq k\leq N$, the shifted quotient
\begin{align*}
\mathring{\bt}_k  (u)
&=
\bQ_k (u-ib/2)\bQ_k  (u+ib/2)^{-1}
\end{align*}
belongs to $\CalT_N[F] $ and satisfies
\begin{align*}
\mathring{\bt}_k(u)
&=
\bigl(L_k(u)\cdots L_1(u)\bigr)_{11}
=
\operatorname{Tr}_{\BC^2}
\left(
\begin{pmatrix}
1&0\\
0&0
\end{pmatrix}
L_k(u)\cdots L_1(u)
\right).
\end{align*}
It is a polynomial of degree $k$,
\begin{align*}
\mathring{\bt}_k(u)
&=
\sum_{r=0}^{k}H_{k,r}e^{2\pi b u r},
\end{align*}
whose coefficients are commuting Hamiltonians,
\[
[H_{k,r},H_{k,s}]=0, 
\qquad
0\leq r,s\leq k.
\]
\end{Theorem}

Here are some Hamiltonians obtained from the transfer matrix
%\begin{subequations}
\begin{gather*}
H_{k,0}=1,
\qquad
H_{k,k}=P_1\cdots P_k,\\
H_{k,1} = \sum_{j=1}^k e^{2\pi bp_j} + \sum_{j=2}^{k} \left(  e^{2\pi b(p_j+x_{j}-x_{j-1})}  + {\bf 1}_F(j) e^{2\pi b(x_{j}-x_{j-1}-m_j)} \right).
\end{gather*}
%\end{subequations}
The theorem identifies the transfer matrix obtained from cluster mutations with the transfer matrix of an open chain, that we call the $q$-deformed flavored Toda chain.
In particular, it gives the local Lax matrices directly from the $Q$-operator construction.
The specialization $m_k \to \infty$ removes the flavor at the $k$-th site, and applying this limit to all $m_k$ recovers the open $q$-deformed Toda chain. 
The commutativity asserted here is an identity in the quantum torus algebra.

The positions of the flavors can be changed by local cluster
transformations.
To state this precisely, we distinguish the positions from the
ordered list of flavor parameters.

\begin{Theorem}[\cref{thm:iso}]
\label{thm:intro-flavor-equivalence}
Let $F=\{f_1<\cdots<f_s\}$, $F'=\{f'_1<\cdots<f'_s\}$ be subsets of $\{2,\dots,N\}$ of the same cardinality.
Assign flavor parameters $m_{f_j}$ and $m'_{f'_j}$ satisfying
\[
m'_{f'_j}=m_{f_j},
\qquad j=1,\dots,s.
\]
There exists a sequence of local cluster transformations inducing an isomorphism together with the canonical transformation,
\[
\Phi:
\operatorname{Frac}\CalT_{N}[F']
\longrightarrow
\operatorname{Frac}\CalT_{N}[F].
\] 
such that
\begin{align*}
\Phi\bigl(\mathring{\bt}'_N(u)\bigr)
&=
\mathring{\bt}_N(u).
\end{align*}
Consequently,
\[
\Phi\bigl(H'_{N,r}\bigr)=H_{N,r},
\qquad r=0,\dots,N.
\]

Each elementary transformation moves a flavor into an adjacent
empty site, preserves its flavor parameter, and fixes the canonical variables
at all other sites.
For such a move between sites $k$ and $k+1$, the transformation
preserves the local Lax product: 
\[
\Phi\bigl(L'_{k+1}(u)L'_k(u)\bigr)
=
L_{k+1}(u)L_k(u).
\]
\end{Theorem}

The theorem shows that the Hamiltonian systems depend, up to
these cluster transformations, on the ordered flavor parameters rather
than their positions along the chain.
The elementary moves exchange a flavor with an empty site and
therefore preserve the relative order of the flavors.

\medskip

We next study the integral operators associated with
$\bQ_N$.
Besides an explicit kernel, we obtain a factorized expression
for the scalar function $Q_{\ba}$ defined by the Fourier
transform in \cref{def:Q-function}.
Writing $\ba=(a_1,\dots,a_{N-1},a_N) \in \mathbb{R}^N$ with $a_N=0$, and $\underline{\ba} = \sum_\alpha a_\alpha$, the result has the form 
\begin{align}\label{eq:intro-scalar-Q}
Q_{\ba}(u)
&=
\CalC_F(\ba,\bm)\,
e^{-2\pi i u \underline{\ba}}
\prod_{\alpha=1}^{N}\varphi(u+a_\alpha),
\end{align}
where $\CalC_F(\ba,\bm)$ is independent of $u$.
Its explicit expression is given in \cref{prop:Q-function-quiver}.
The Fourier calculation is understood in the meromorphic sense
specified by the integral identities and contour prescriptions
used there.
In particular, the normalized expression
\[
\widehat{Q}_\ba (u) := \frac{
e^{2\pi i u \underline{\ba}}
Q_{\ba}(u)}
{\CalC_F(\ba,\bm)}
=
\prod_{\alpha=1}^{N}\varphi(u+a_\alpha)
\]
is independent of the flavor positions and the associated parameters.

This factorization suggests a connection with separation of
variables. The functional equation of the quantum dilogarithm gives
\[
\widehat{Q}_{\ba}(u-ib/2)
=
\left[
\prod_{\alpha=1}^{N}
\left(1+e^{2\pi b(u+a_\alpha)}\right)
\right]
\widehat{Q}_{\ba}(u+ib/2),
\]
which is a scalar difference equation of Baxter form.
Moreover, $\widehat{Q}_{\ba}(u)$ agrees with the expression
for the eigenvalue of the unflavored Baxter operator on
the $b$-Whittaker functions
\cite[Proposition~8.3]{schrader2018b}.
In the separation of variables approach, scalar Baxter
functions provide building blocks for common
eigenfunctions of the commuting Hamiltonians \cite{Lee:2020hfu}.
These observations motivate investigating a corresponding
construction for the flavored chains.
Such a construction would require an explicit separation of variable transform and an identification of the parameters $\ba$
with joint spectral data, which are beyond the scope of
the present paper.

The same factorization motivates the gauge-theoretic part of the paper.
Five-dimensional $\CalN=1$ theories engineered by toric
Calabi--Yau threefolds admit BPS-quiver descriptions after
compactification on a circle~\cite{Closset:2019juk}.
We relate the quivers $\CalQ_N[F]$ to periodic BPS quivers
through a reduction associated with the bulk-decoupling limit.
We then consider a canonical codimension-two defect obtained by
Higgsing a higher-rank gauge theory.
On a squashed five-sphere, this defect wraps a squashed
three-sphere, and its perturbative contribution is expressed
through double sine functions.
Their expression in terms of quantum dilogarithms allows a
comparison with~\eqref{eq:intro-scalar-Q}, including the
elementary exponential factors and normalization choices.
This comparison concerns the perturbative defect integrand.
A physical interpretation of the distinguished mutation sequence,
and an extension to finite bulk coupling, remain to be understood.

%\medskip
\subsection*{Organization of the paper}

The paper is organized as follows.
\cref{sec:qcm} reviews quantum cluster mutations and fixes
our conventions.
\cref{sec:Baxter Q-operator} introduces the quivers
$\CalQ_N[F]$, constructs the $Q$-operators, and derives their
integral kernels and Fourier transforms.
\cref{sec:Integrable} proves the commutativity statements,
establishes the Lax realization, and constructs the local
transformations moving flavors.
\cref{sec:Example} gives explicit examples for $N=3$ and
$N=4$.
\cref{sec:SUSY} discusses the relation to five-dimensional
BPS quivers and canonical surface defects.
\cref{sec:functions} records the special-function
conventions and identities used throughout the paper.

\subsection*{Acknowledgments}

The work of SJ is supported by the Institute for Basic Science under the project IBS-R003-Y3. The work of TK is supported by EIPHI Graduate School (No.~ANR-17-EURE-0002) and the Bourgogne-Franche-Comté region.
The work of NL is supported by the Institute for Basic Science under the project IBS-R003-D1.

\section{Quantum cluster mutation}\label{sec:qcm}

In this section we review the basics of quantum cluster mutations following \cite{fock2009cluster}. 
A cluster seed $\Theta$ is a datum $\Theta=(\L,(\cdot,\cdot),\{e_i\})$ where
\begin{itemize}
    \item $\L$ is a finite-dimensional lattice.
    \item $(\cdot,\cdot)$ is a $\BZ$-valued skew-symmetric form on $\L$.  
    \item $\{e_i \mid i \in I\}$ is a set of basis of the lattice $\L$. 
\end{itemize}
We assume that all nodes are mutable.

We associate each seed $\T$ to a quiver $\CalQ$ whose vertices $V=\{v_i\}_{i\in I}$ are labeled by the set $I$ and oriented arrows specified by an adjacency matrix $\ve=(\ve_{ij})_{i,j \in I}$ with $\ve_{ij} = (e_i,e_j)$. 
The \emph{mutation of $\T$ on the direction $k$}, denoted by $\m_k(\T)$, is defined by change of the basis $\mu_k : \{e_j\} \mapsto \{e'_j\}$ where 
\begin{align}\label{def:mutated_basis}
    e'_j = \begin{cases}
        -e_j & \text{if } j=k \\
        e_j + \max(\ve_{jk},0) e_k & \text{if } j\neq k
    \end{cases}.
\end{align}
At the level of the quiver, the mutated quiver $\CalQ'$ is obtained from $\CalQ$ by the following steps: 
\begin{enumerate}
    \item For every oriented 2-step path going through $v_k$, i.e., $v_j \to v_k \to v_i$, add an arrow from $v_j \to v_i$. 
    \item Reverse the direction of every single arrows whose source or target are $v_k$. 
    \item Remove any 2-cycle, a pair of arrows $v_i \substack{\to \\ \gets} v_j$, created in the first step between two vertices $v_i$ and $v_j$. 
\end{enumerate}
The adjacency matrix $\ve'=(\ve'_{ij})$ for the mutated quiver $\CalQ'$ is given by
\begin{align}\label{def:mutated_edges}
    \ve_{ij}' = 
    \begin{cases}
        -\ve_{ij} & \text{if $i=k$ or $j=k$} \\
        \ve_{ij} + \frac{|\ve_{ik}|\ve_{kj} + \ve_{ik}|\ve_{kj}|}{2} & \text{otherwise}
    \end{cases}.
\end{align}

\begin{Definition}[Quiver isomorphism]\label{def:isomorphic}
    Let $\CalQ$ and $\CalQ'$ be two quivers. We say $\CalQ'$ is isomorphic to $\CalQ$ if
    there exists a bijective map between the index sets of the vertices $c:I \to I'$ such that 
    \[
        \ve'_{c(i),c(j)} = \ve_{ij}. 
    \]
\end{Definition}

We would like to remark that $\T$ and $\m_k^2(\T)$ share the same adjacency matrix $\ve$, hence the associated quivers are isomorphic by \cref{def:isomorphic}. But their basis are not necessarily identical.

The seed $(\L,(\cdot,\cdot))$ determines a \emph{quantum torus algebra} $\CalT_\L$, which is a free $\BZ[q^\pm]$-module, $q\in \BC^\times$, spanned by the \emph{cluster variables} $X_\l$, $\l\in \L$, with $X_0 = 1$ and the multiplicity given by 
\begin{align*}
    q^{(\l,\m)} X_\l X_\m = X_{\l+\m}~,~ \l,\m\in \L. 
\end{align*}
The generators $X_{e_i}$ of $\CalT_\L$ correspond to basis elements $e_i$ of $\L$. 

Given $\CalT_\L$, one can associate a Heisenberg $\ast$-algebra $\CalH_\L$. We abuse the notation and denote the algebra's generators by the same symbol $\{v_i\}$ as the vertices of the quiver $\CalQ$. It is a topological $\ast$-algebra over $\BC$ generated by elements $\{v_i\}_{i\in I}$  satisfying 
\begin{align*}
    [v_j,v_k] = \frac{1}{2\pi i} \ve_{jk}~,~ \ast v_j = v_j~.
\end{align*}
The assignment 
\begin{align*}
    X_{e_j} = e^{2\pi b v_j}~,~ q=e^{\pi i b^2}
\end{align*}
with $b\in\BR_{>0}$
defines a homomorphism from $\CalT_\L$ to $\CalH_\L$. 

For each mutation $\m_k$ we define an algebra automorphism of the skew field denoted by $\operatorname{Frac}\CalT_\L$ 
\begin{align*}
    \m_k(X_{e_j}) = 
    \begin{cases}
        X_{e_j'} \left( 1 + q X_{e_k} \right) & \text{if } \ve_{ki} = 1, \\
        X_{e_j'} (1+qX_{e_k}^{-1})^{-1} & \text{if } \ve_{ki} = -1. 
    \end{cases}
\end{align*}
where $\{e_i'\}$ is the basis of the mutated seed $\m_k(\T)$ given in \eqref{def:mutated_basis}.
By the homomorphism from $\CalT_\L$ to $\CalH_\L$, such mutation can be written as an adjoint operator in the Heisenberg $\ast$-algebra. We will abuse the notation here and call this automorphism a \emph{quantum mutation}. 
Using the properties of the quantum dilogarithm function $\varphi$ shown in \eqref{eq:qdf-conjugation}, the mutation is expressed as an adjoint operator acting on the Heisenberg $\ast$-algebra $\CalH_\L$:
\begin{align}\label{eq:mutation-adjoint}
    \m_k = \Ad_{\varphi(-v_k)^{-1}} \in \operatorname{End}(\CalH_\L).
\end{align}

\section{Cluster realization of quantum Baxter Q-operator}\label{sec:Baxter Q-operator}

\begin{Definition}\label{def:quiver}
    For $N \in \mathbb{Z}_{>1}$, we define a quiver $\CalQ_{N}$ with $3N-1$ vertices $V=\{v_i\}_{i=0,\ldots,3N-2}$, and the following set of arrows: 
    \if0
    \begin{itemize}
        \item For all $1\leq k\leq N-1$, there is an arrow from $v_{3k} \to v_{3k-1}$, an arrow from $v_{3k} \to v_{3k-2}$, an arrow from $v_{3k-1} \to v_{3k-2}$. 
        \item For all $1\leq k \leq N$, there is an arrow from $v_{3k-2} \to v_{3k-3}$. 
        \item For all $1\leq k\leq N-2$, there is an arrow from $v_{3k-2} \to v_{3k+1}$, an arrow from $v_{3k-1} \to v_{3k+3}$, an arrow from $v_{3k+1} \to v_{3k-2}$. 
        \item There is an arrow from $v_0 \to v_3$, an arrow from $v_{3N-5} \to v_{3N-2}$, an arrow from $v_{3N-2} \to v_{3N-3}$. 
    \end{itemize}
    \fi
    \[
    A = \left\{ \substack{\displaystyle \{v_{3k} \to v_{3k-1}, v_{3k} \to v_{3k-2}, v_{3k-1} \to v_{3k-2}\}_{k=1}^{N-1}, \ \{v_{3k-2} \to v_{3k-3} \}_{k=1}^N, \\ \displaystyle \{v_{3k-2} \to v_{3k+1}, v_{3k-1} \to v_{3k+3}, v_{3k+1} \to v_{3k-2} \}_{k=1}^{N-2}, \\ \displaystyle  v_0 \to v_3, \ v_{3N-5} \to v_{3N-2}, \ v_{3N-2} \to v_{3N-3} }\right\} .
    \]
    Two vertices $v_0$ and $v_{3N-2}$ in $\CalQ_N$ are connected by only two arrows, which we call \emph{end vertices} or \emph{handles}. 
    We denote the quantum torus algebra associated with $\CalQ_N$ by $\CalT_N$. 
\end{Definition}

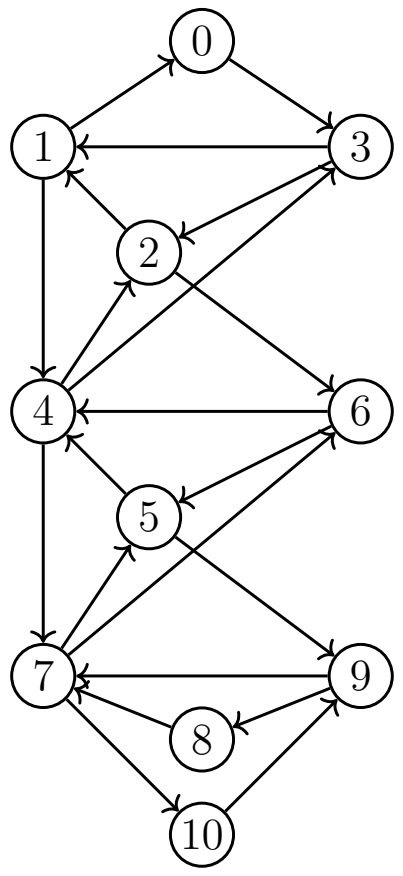
\begin{figure}[h!]
% \if0
\centering
\scalebox{.7}{
\begin{tikzpicture}[square/.style={regular polygon,regular polygon sides=4},decoration={markings, 
    mark= at position 0.5 with {\arrow[scale=1.3]{stealth}}}
 ]

    %%% Quiver 1 %%%%%%
    
    % \node at (-2.5,-2) [style={circle,draw},thick,inner sep=6pt] (r0) {};
    % \node at (-4,-3) [style={circle,draw},thick,inner sep=6pt] (r1) {};
    % \node at (-3,-4) [style={circle,draw},thick,inner sep=6pt] (r3) {};
    % \node at (-1,-3) [style={circle,draw},thick,inner sep=6pt] (r2) {};
    % \node at (-4,-5.5) [style={circle,draw},thick,inner sep=6pt] (r4) {};
    % \node at (-1,-5.5) [style={circle,draw},thick,inner sep=6pt] (r5) {};
    % \node at (-2.5,-6) [style={circle,draw},thick,inner sep=6pt] (r6) {};
    % \node at (-2.5,-7) [style={circle,draw},thick,inner sep=6pt] (r7) {};

    % % \node at (-2.5,-.7) (Q1){$\CalQ_3$};
    % \node at (-2.5,-2) {0};
    % \node at (-4,-3) {1};
    % \node at (-1,-3) {3};
    % \node at (-3,-4) {2};
    % \node at (-4,-5.5) {4};
    % \node at (-1,-5.5) {6};
    % \node at (-2.5,-6) {5};
    % \node at (-2.5,-7) {7};

    % %%% Arrows %%%

    % \draw[thick,->] (r1) -- (r0);
    % \draw[thick,->] (r0) -- (r2);
    % \draw[thick,->] (r2) -- (r1);
    % \draw[thick,->] (r1) -- (r4);
    % \draw[thick,->] (r4) -- (r2); 
    % \draw[thick,->] (r2) -- (r3);
    % \draw[thick,->] (r3) -- (r1);
    % \draw[thick,->] (r4) -- (r3);
    % \draw[thick,->] (r3) -- (r5);
    % \draw[thick,->] (r5) -- (r4); 
    % \draw[thick,->] (r4) -- (r7);
    % \draw[thick,->] (r6) -- (r4);
    % \draw[thick,->] (r5) -- (r6);
    % \draw[thick,->] (r7) -- (r5);

    %%%%% QUIVER %%%%%
    
    \node at (-12.5,-2) [style={circle,draw},thick,inner sep=6pt] (r0) {};
    \node at (-14,-3) [style={circle,draw},thick,inner sep=6pt] (r1) {};
    \node at (-13,-4) [style={circle,draw},thick,inner sep=6pt] (r3) {};
    \node at (-11,-3) [style={circle,draw},thick,inner sep=6pt] (r2) {};
    \node at (-14,-5.5) [style={circle,draw},thick,inner sep=6pt] (r4) {};
    \node at (-11,-5.5) [style={circle,draw},thick,inner sep=6pt] (r5) {};
    \node at (-13,-6.5) [style={circle,draw},thick,inner sep=6pt] (r6) {};
    \node at (-14,-8) [style={circle,draw},thick,inner sep=6pt] (r7) {};
    \node at (-11,-8) [style={circle,draw},thick,inner sep=6pt] (r8) {};
    \node at (-12.5,-8.6) [style={circle,draw},thick,inner sep=6pt] (r9) {};
    \node at (-12.5,-9.5) [style={circle,draw},thick,inner sep=6pt] (r10) {};

    % \node at (-12.5,-1.2) (Q1){$\mathcal{Q}_4$};
    \node at (-12.5,-2) {0};
    \node at (-14,-3) {1};
    \node at (-11,-3) {3};
    \node at (-13,-4) {2};
    \node at (-14,-5.5) {4};
    \node at (-11,-5.5) {6};
    \node at (-13,-6.5) {5};
    \node at (-14,-8) {7};
    \node at (-11,-8) {9};
    \node at (-12.5,-8.6) {8};
    \node at (-12.5,-9.5) {10};

    %%%%% Arrow %%%%%%

    \draw[thick,->] (r0) -- (r2);
    \draw[thick,->] (r1) -- (r0);
    \draw[thick,->] (r1) -- (r4);
    \draw[thick,->] (r2) -- (r1);
    \draw[thick,->] (r2) -- (r3);
    \draw[thick,->] (r3) -- (r1);
    \draw[thick,->] (r3) -- (r5);
    \draw[thick,->] (r4) -- (r2);
    \draw[thick,->] (r4) -- (r3);
    \draw[thick,->] (r4) -- (r7);
    \draw[thick,->] (r5) -- (r4);
    \draw[thick,->] (r5) -- (r6);
    \draw[thick,->] (r6) -- (r4);
    \draw[thick,->] (r6) -- (r8);
    \draw[thick,->] (r7) -- (r5);
    \draw[thick,->] (r7) -- (r6);
    \draw[thick,->] (r8) -- (r9);
    \draw[thick,->] (r8) -- (r7);
    \draw[thick,->] (r9) -- (r7);

    \draw[thick,->] (r7) -- (r10);
    \draw[thick,->] (r10) -- (r8);

\end{tikzpicture}
}
% \fi
\caption{Quiver $\CalQ_4$.} 
\label{fig:Q3}
\end{figure}

See \cref{fig:Q3} for the case $N=4$.

Consider the quiver $\CalQ_N$ and the corresponding cluster seed together with the associated Heisenberg $\ast$-algebra $\CalH_N$. 
Let $\bm=\{m_k\}_{k=2}^{N}\in\BR^{N-1}$ the collection of parameters, that we call the \emph{flavor parameters}.%
\footnote{This terminology comes from its relation to the underlying supersymmetric gauge theory. See \cref{sec:SUSY}.}
For any $u,w \in \mathbb{R}$, assignment 
\begin{align}\label{eq:cluster_to_canonical}
\begin{split}
    & v_0 = -\hat{p}_1-u, \quad  
    v_{3k-2} = \hat{x}_{k+1}-\hat{x}_k+\hat{p}_k +m_{k+1}, \\
    & v_{3k-1} = -\hat{p}_{k+1}-m_{k+1}, \quad  
    v_{3k} = \hat{x}_{k}-\hat{x}_{k+1}, \quad 
    v_{3N+1} = \hat{p}_{N} + w,
\end{split}
    \qquad 
    k=1,\dots,N-1, 
\end{align}
defines a representation of $\CalH_N$ on $L^2(\mathbb{R}^N)$ in the variables $\bx=(x_1,\dots,x_N)\in \BR^N$, 
in which the generator $\hat{x}_j$ acts by unbounded self-adjoint operator. Namely, for $f$ belongs to the representation module of $\mathcal{H}_N$, we have 
\[
    \hat{x}_j f(\bx) = x_j f(\bx).
\]
We define a self-adjoint operator on $\mathcal{H}_N$,
\[
    \hat{p}_j f(\bx) = \frac{1}{2\pi i} \frac{\p}{\p x_j} f(\bx), 
\]
which is the conjugate momentum of $\hat{x}_k$ obeying
\[
    [\hat{p}_j,\hat{x}_k] = \frac{\d_{jk}}{2\pi i}. 
\]
We will omit the hat sympol from now on. 

\begin{Definition}[Merging vertices]\label{def:merging}
    Given a quiver $\CalQ$ with vertices $V_\CalQ=\{\dots,v_j,\dots,v_k,\dots\}$, merging $v_j$ to $v_k$ is defined to generate a new quiver $\CalQ'$ with vertices $V_{\CalQ'}=V_\CalQ \cup \{v_j'\} \backslash\{v_j,v_k\}$.  The adjacency matrix and generators of Heisenberg $\ast$-algebra for the vertex $v'_j$ in $\CalQ'$ are defined as
    \begin{align*}
    \begin{split}
        & \ve_{j,l}' = \ve_{j,l} + \ve_{k,l}, \quad l\neq j,k, \\ 
        & v_j'=v_j+v_k~. \\
    \end{split}
    \end{align*}
\end{Definition}
Namely, when two vertices $e_j$ and $e_k$ merge in a quiver $\CalQ$, the merged vertex $e_j'$ inherits all arrows that connect to the original two vertices. 

\begin{Definition}\label{def:QFN}
    For $N\geq 2$ and a collection of positions $F\subseteq\{2,\dots,N\}$, we define a quiver $\CalQ_{N}[F]$ by merging $v_{3k-4}$ to $v_{3k-5}$ in $\CalQ_N=\CalQ_{N}[\{2,\dots,N\}]$ if $k \notin F$. 
    We denote the quantum torus algebra associated with $\CalQ_N[F]$ by $\CalT_N[F]$. 
\end{Definition}
The generators $\{v_i\}$ of the Heisenberg $\ast$-algebra associated with $\CalQ_N[F]$ are given by
\begin{align*}
\begin{split}
    & v_0 = -p_1-u, \quad 
    v_{3k-2} = \begin{cases}
        x_{k+1}-x_k+p_k +m_{k+1} & \text{if } k+1 \in F \\
        x_{k+1}-x_k+p_k-p_{k+1} & \text{if } k+1 \notin F 
    \end{cases},~ \\
    & v_{3k-1} = \begin{cases}
        -p_{k+1}-m_{k+1} & \text{if } k+1\in F \\
        \text{n/a} & \text{if } k+1 \notin F
    \end{cases}, \\
    & v_{3k} = x_{k}-x_{k+1}, \quad 
    v_{3N+1} = p_{N} + v.
\end{split}
\end{align*}
for $k=1,\dots,N-1$. 

Given a quiver $\CalQ$, suppose there are arrows $v_i \to v_j$, $v_j \to v_k$, and $v_k \to v_i$ in $\CalQ$. 
We denote such arrows as a 3-cycles by 
\begin{align}\nonumber
\begin{split}
    & [i~j~k]:v_i \to v_j \to v_k \to v_i, \quad
    [i~j~k] = [j~k~i] = [k~i~j].
\end{split}
\end{align}

\begin{Definition}
Let $\D_{i_1,i_2,i_3}=(i_1,i_2,i_3)$ be a triangle in $\CalQ_N$ whose vertices are $v_{i_1}$, $v_{i_2}$, and $v_{i_3}$ such that
\[
    \ve_{i_2,i_1} = \ve_{i_3,i_1} = \ve_{i_3,i_2} = 1.
\]
For each triangle $\D_{i_1,i_2,i_3}$, we consider 3-cycles sharing its edges
\begin{align}\nonumber
\begin{split}
    \D_{i_1,i_2,i_3}^+ =\{[i_1~i_4~i_2]\mid \Delta_{i_1,i_2,i_3} = (i_1,i_2,i_3),~ i_4 \in I \}, \\
    \D_{i_1,i_2,i_3}^- =\{[i_2~i_4~i_3]\mid \Delta_{i_1,i_2,i_3} = (i_1,i_2,i_3),~ i_4 \in I \}.
\end{split}
\end{align}
We define the orientation $d_{i_1,i_2,i_3}$ of the triangle $\D_{i_1,i_2,i_3}$ by
\begin{align*}
    d_{i_1,i_2,i_3} := |\D_{i_1,i_2,i_3}^+| - |\D_{i_1,i_2,i_3}^-|. 
\end{align*}
\end{Definition}

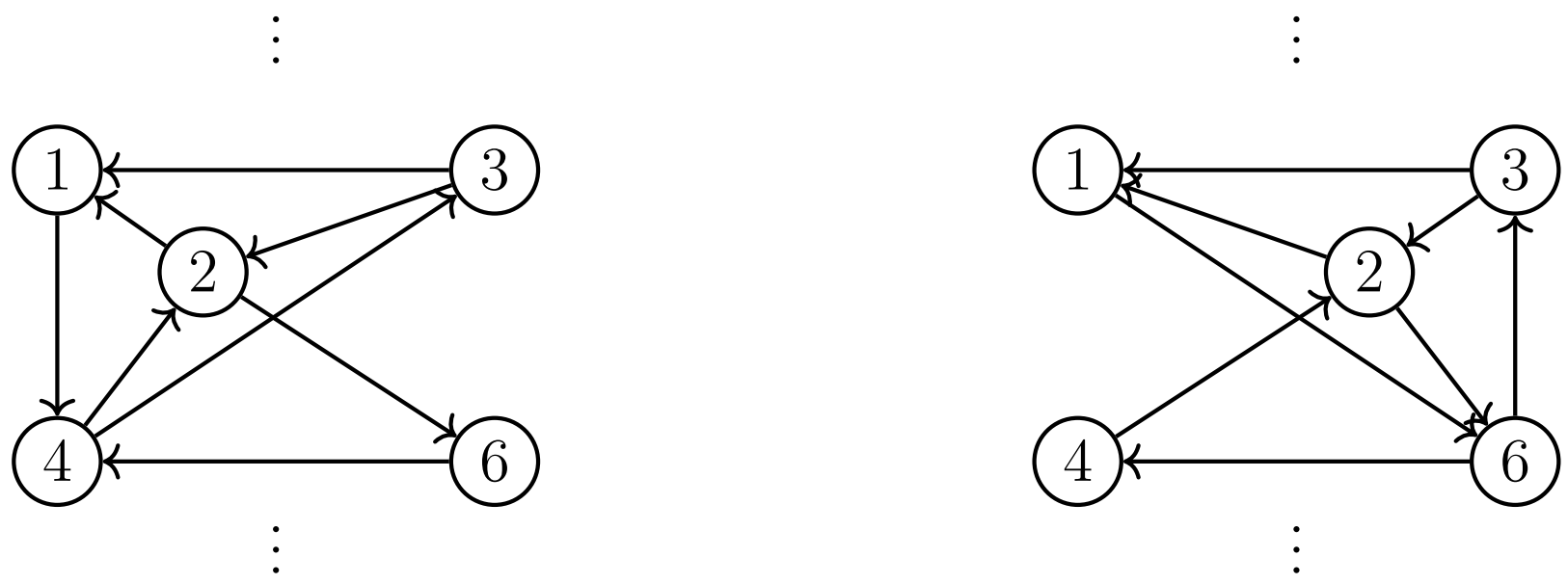
\begin{figure}[h] 
% \if0
\centering
\scalebox{1}{
\begin{tikzpicture}[square/.style={regular polygon,regular polygon sides=4},decoration={markings, 
    mark= at position 0.5 with {\arrow[scale=1.3]{stealth}}}
 ]

    %%% ORIGINAL %%%
    
    %%% Node %%%

    % \node at (-7.5,-0) [fill,circle,green,inner sep=6pt] (q0) {};
    \node at (-9,-1) [style={circle,draw},thick,inner sep=6pt] (q1) {};
    \node at (-8,-1.7) [style={circle,draw},thick,inner sep=6pt] (q2) {};
    \node at (-6,-1) [style={circle,draw},thick,inner sep=6pt] (q3) {};
    \node at (-9,-3) [style={circle,draw},thick,inner sep=6pt] (q4) {};
    \node at (-6,-3) [style={circle,draw},thick,inner sep=6pt] (q5) {};

    \node at (-9,-1) {1};
    \node at (-8,-1.7) {2};
    \node at (-6,-1) {3};
    \node at (-9,-3) {4};
    \node at (-6,-3) {6};
    % \node at (q0) {0};
    % \node at (q6) {7};

    \node at (-7.5,0) {$\vdots$};
    \node at (-7.5,-3.5) {$\vdots$};
    \node at (-7.5,-4.5) {$\D_{1,2,3}^+=\{[1~4~2]\}$, $\D^-_{1,2,3}=\varnothing$.}; 
    \node at (-7.5,-5) {$d_{123}=1$.}; 
    
    %%% Arrows %%%

    % \draw[thick,->] (q1) -- (q0);
    % \draw[thick,->] (q0) -- (q3);
    
    \draw[thick,->] (q5) -- (q4);
    \draw[thick,->] (q1) -- (q4);
    \draw[thick,->] (q2) -- (q1); 
    \draw[thick,->] (q2) -- (q5);
    \draw[thick,->] (q3) -- (q1);
    \draw[thick,->] (q3) -- (q2);
    \draw[thick,->] (q4) -- (q2);
    \draw[thick,->] (q4) -- (q3);

    % \draw[thick,->] (q4) -- (q6);
    % \draw[thick,->] (q6) -- (q5);

    %%% REFLECTION %%%

    %%% Node %%%

    % \node at (-1.5,-0) [fill,circle,green,inner sep=6pt] (p0) {};
    \node at (-2,-1) [style={circle,draw},thick,inner sep=6pt] (p1) {};
    \node at (-0,-1.7) [style={circle,draw},thick,inner sep=6pt] (p2) {};
    \node at (1,-1) [style={circle,draw},thick,inner sep=6pt] (p3) {};
    \node at (-2,-3) [style={circle,draw},thick,inner sep=6pt] (p4) {};
    \node at (1,-3) [style={circle,draw},thick,inner sep=6pt] (p5) {};
    % \node at (-1.5,-4) [fill,circle,red,inner sep=6pt] (p6) {};

    \node at (-2,-1) {1};
    \node at (-0,-1.7) {2};
    \node at (1,-1) {3};
    \node at (-2,-3) {4};
    \node at (1,-3) {6};
    % \node at (p0) {0};
    % \node at (p6) {7};

    \node at (-.5,0) {$\vdots$};
    \node at (-.5,-3.5) {$\vdots$};
    \node at (-.5,-4.5) {$\D^+_{1,2,3}=\varnothing$, $\D_{1,2,3}^-=\{[2~6~3]\}$}; 
    \node at (-.5,-5) {$d_{1,2,3}=-1$}; 
    
    %%% Arrows %%%

    % \draw[thick,->] (p1) -- (p0);
    % \draw[thick,->] (p0) -- (p3);

    \draw[thick,->] (p5) -- (p4);
    \draw[thick,->] (p1) -- (p5);
    \draw[thick,->] (p2) -- (p1);
    \draw[thick,->] (p2) -- (p5);
    \draw[thick,->] (p3) -- (p1);
    \draw[thick,->] (p3) -- (p2);
    \draw[thick,->] (p4) -- (p2);
    \draw[thick,->] (p5) -- (p3);
    % \draw[thick,->] (p5) -- (p4);

    % \draw[thick,->] (p4) -- (p6);
    % \draw[thick,->] (p6) -- (p5);

 \end{tikzpicture}
}
% \fi
\caption{The triangle $\D_{123}$ in a quiver $\CalQ$. On the left the triangle has positive orientation. On the right the triangle has negative orientation.} 
\label{fig:orientate_trio}
\end{figure}

See \cref{fig:orientate_trio} for an illustration. 
We remark that for $\CalQ_N$ there are $N-1$ such triangles
\[
    \D_{3k-5,3k-4,3k-3} , \quad k=2,\dots,N. 
\]
All the triangles except the last one have positive orientations 
\[
    d_{3k-5,3k-4,3k-3}=1-\d_{k,N}, \quad k =2,\dots,N. 
\] 
We say that these triangles $\D_{3k-5,3k-4,3k-3}$ \emph{representing flavors}. By \cref{def:QFN}, $\CalQ_N[F]$ has exactly $|F|$ triangles $\{\D_{3k-5,3k-4,3k-3}\}_{k \in F}$.  

\begin{Definition}[Baxter $\bQ$-operator] \label{def:Baxter-Q}
    We consider the following mutation sequence for $\CalQ_N$:
    \[
        \m^{(1)} = \m_{3N-3} \m_{3N-4} \cdots \m_1 \m_0 .
    \]
    We define the \emph{Baxter $\bQ$-operator} as the inverse of $\mu^{(1)}$ acting on $\mathcal{H}_N$,
    \begin{align*}
        \bQ_N(u)^{-1} = \m^{(1)} = \m_{3N-3} \cdots \m_1\m_0 \in \operatorname{End}(\CalH_N).
    \end{align*}
    For $\CalQ_N[F]$ with fewer flavors $F\subseteq\{2,\dots,N\}$ the Baxter $\bQ$-operator given by taking
    \[
        \m_{3k-4} = 1 \text{ if } k\notin F. 
    \]
\end{Definition}

\begin{Remark}[Follow the arrow rule]\label{def:follow the arrow}
    The mutation sequence $\m^{(1)}$ is given by the \emph{follow the arrow rule}: 
    \begin{enumerate}
        \item Start with a mutation on the vertex $v_0$. After mutation $\mu_0$, there will be one unique arrow pointing toward another vertex $v_1$.
        \item Follow the arrow and mutate the vertex that the arrow pointing towards, i.e. $\mu_1$. After mutation there should be two arrows pointing outwardly from the mutated vertex, one pointing forward to $v_2$ and the other pointing back to the previous vertex $v_0$. 
        \item Follow the arrow and repeat process 2. 
        \item Up to some point (\# of mutation = \# of nodes in the quiver $-1$) we reach the last vertex $v_{3N-3}$ where there will be only one arrow pointing backward to $v_{3N-4}$ after mutation. This completes the first \emph{follow the arrow} mutation sequence.     
    \end{enumerate}
\end{Remark}

Using \eqref{eq:cluster_to_canonical}, the Baxter $\bQ$-operator now acts on the representation of $\CalH_N$ on the Hilbert space of $L^2$-functions in  $\bx=(x_1,\dots,x_N)$, whose conjugate momenta are labeled by $\mathbf{p}=(p_1,\dots,p_N)$. 

\begin{Proposition}\label{prop:Baxter-Q}
    For $\CalQ_{N}[F]$, the Baxter $\mathbf{Q}$-operator is given by
    \begin{align}\nonumber
        \bQ_1=\varphi(p_1+u)~,~ \bQ_N(u)= \bQ_1(u) T_2(u) \cdots T_{N}(u)
    \end{align}
    where
    \begin{align}\nonumber
        T_k(u) = \varphi(x_k-x_{k-1}+u-m_k)^{\mathbf{1}_F(k)}\varphi(p_k+x_k-x_{k-1}+u)\varphi(p_k+u)
    \end{align}
    with the indicator function
    \begin{align}\label{eq:true-or-false}
        \mathbf{1}_F(k) = \begin{cases}
            1 & \text{if } k \in F, \\
            0 & \text{if } k \notin F.
        \end{cases} 
    \end{align}
\end{Proposition}
\begin{proof}
    For $\mathcal{Q}_N$, the result immediately follows by \eqref{eq:mutation-adjoint}. 
    For $\mathcal{Q}_N[F]$ in general, we take the limit $m_k \to \infty$ for $k \not\in F$ to obtain the result.
\end{proof}

In the absence of any flavors $F=\varnothing$, our construction of the Baxter $\bQ$-operator is identical to that of \cite{schrader2018b}.

Let $\CalQ_N^{(1)} = \m^{(1)} \CalQ_N$. 
We denote the set of vertexes of $\CalQ_N^{(1)}$ by $V^{(1)}=\{v_1^{(1)},\dots,v_{3N-1}^{(1)}\}$ where
\[
    v_{j}^{(1)} = v_{j-1}~,~ j=1,\dots,3N-1.
\]
Compared with $\CalQ_N$, $\CalQ_N^{(1)}$ is modified as follows:
\begin{enumerate}
    \item The vertex $v_0$ is removed. 
    \item There is an additional vertex $v^{(1)}_{3N-1}$ with arrows, $v^{(1)}_{3N-5} \to v_{3N-1}^{(1)}$ and $v_{3N-1}^{(1)} \to v_{3N-3}^{(1)}$.
\end{enumerate}
At the level of the quiver, the effect of the $\m^{(1)}$ mutation sequence is to transport the ``handle'' vertex $v_0$ from top to bottom of the quiver.

We then consider a mutation sequence that changes orientation of the triangles in $\CalQ_N^{(1)}$. 
We define the sequence of mutations
    \[
        \s_k := \m_{3k-2} \m_{3k-1} \cdots \m_{3N-8}\m_{3N-7} \m_{3N-5} \m_{3N-4}~, \quad \quad k=2,\dots, N-1,
 \]
which changes the orientation of the $(k-1)$-th triangle $\D_{3k-5,3k-4,3k-3}$ in $\CalQ_N^{(1)}$ with a cyclic permutation of the vertices 
    \[
        (v_{3j-3},v_{3j-2},v_{3j-1}) \mapsto (v_{3j-2},v_{3j-1},v_{3j-3}) , \quad 
        k\leq j\leq N-1 .
    \]
Let $\eta$ be a cyclic permutation operator 
    \[
       \eta : (v_{3k-3},v_{3k-2},v_{3k-1}) \mapsto (v_{3k-2},v_{3k-1},v_{3k-3})~, \quad k=1,\dots,N-1, 
    \]
and we denote 
\[
    \s_k^{(n)} : = \m_{\eta^n(3k-2)}\m_{\eta^n(3k-1)} \cdots \m_{\eta^{n}(3N-5)} \m_{\eta^n(3N-4)} .
\]

\begin{Proposition}\label{def:mu^{(2)}} 
The mutation sequence 
    \begin{align*}
        \m^{(2)} = \s_{N-1}^{(N-3)}\cdots \s_3^{(1)}\s_2.
    \end{align*}
    changes orientation of all the triangles in $\CalQ_N^{(1)}$. 
\end{Proposition}
\begin{proof}
    The consecutive mutation of $\s_k$ changes the orientation of the triangle $\D_{3k-5,3k-4,3k-3}$ while permuting the vertices below. $\s^{(k-2)}_{k}$ is to keep track of permutation of the vertices so that the consecutive mutation still change the orientation of triangles. 
\end{proof}

We remark here that $\m^{(2)}$ given in \cref{def:mu^{(2)}} is one of the sequences of mutations that change orientation of all the triangles in $\CalQ_N^{(1)}$. It may be neither the only nor the shortest sequence of mutations to achieve change of orientations. 
The case of $\CalQ_N[F]^{(1)}$ with $|F| < N-1$ can be reduced from $\CalQ_N^{(1)}$. By \cref{def:QFN}, we set $\mu_{3k-4} = 1$ if $k \notin F$. 
If $(k-1) \in F$, i.e. there exist a triangle $\D_{3k-5,3k-4,3k-3}$ in $\CalQ_N[F]^{(1)}$, the sequence of mutation $\s_k$ will flip the orientation of the triangle $\D_{3k-5,3k-4,3k-3}$ and permutate the vertices 
\begin{align}\nonumber
    \begin{cases}
        (v_{3j-3},v_{3j-2},v_{3j-1}) \mapsto (v_{3j-2},v_{3j-1},v_{3j-3}) & j+1 \in F \\
        (v_{3j-3},v_{3j-1}) \mapsto (v_{3j-1},v_{3j-3}) & j+1 \notin F
    \end{cases}
\end{align}
for all $k \leq j \leq N-1$.
Let $\eta_F$ be a cyclic permutation operator
    \[
       \eta_F : 
       \begin{cases}
        (v_{3k-3},v_{3k-2},v_{3k-1}) \mapsto (v_{3k-2},v_{3k-1},v_{3k-3}) & k+1 \in F \\
        (v_{3k-3},v_{3k-1}) \mapsto (v_{3k-1},v_{3k-3}) & k+1 \notin F
    \end{cases}
    \]
We denote 
\[
    \s_k^{(n)} : = \m_{\eta^n_F(3k-2)}\m_{\eta^n_F(3k-1)} \cdots \m_{\eta^{n}_F(3N-5)} \m_{\eta^n_F(3N-4)}.
\]
The $\m^{(2)}$ for $\CalQ_N[F]^{(1)}$ is then defined the same as \cref{def:mu^{(2)}}. 

\begin{Definition}\label{def:second-follow-the-arrow}
    Let $\CalQ_N^{(2)} = \mu^{(2)} \CalQ_N^{(1)}$.
    We define the following mutation sequence for $\CalQ_N^{(2)}$ by
    \begin{align*}
    \begin{split}
    \m^{(3)} = &~ \m_2 \m_1 \m_0 \cdots \m_{\eta^{k-2}(3k-4)}\m_{\eta^{k-2}(3k-5)} \m_{\eta^{k-2}(3k-6)}  \\
    & ~\cdots \m_{\eta^{N-2}(3N-4)}\m_{\eta^{N-2}(3N-5)} \m_{\eta^{N-2}(3N-6)} \m_{3N-2} .
    \end{split}
    \end{align*}
For the case with fewer flavors $F\subset\{2,\dots,N\}$, we take $\m_{3k-4} = 1$ if $k \notin F$ and change $\eta \to \eta_F$ in the definition of $\m^{(3)}$. 
\end{Definition}
The sequence of mutations $\m^{(3)}$ also has an interpretation as the \emph{follow the arrow rule} shown in \cref{def:follow the arrow} with the starting vertex being $v_{3N-2}$. 

\begin{Proposition}\label{prop:swap} 
    Let $\CalQ_N^{(3)}[F] = ( \mu^{(3)} \circ \mu^{(2)} \circ \mu^{(1)} ) \CalQ_N[F]$.
    Change the orientation of all the triangles in $\CalQ_N^{(3)}[F]$.
    Then, the resulting quiver denoted by $\CalQ_N^{(4)}[F]$ is isomorphic to the original quiver $\CalQ_N[F]$.
\end{Proposition}

\begin{proof}
The mutation sequence modifies the quiver $\CalQ_N[F]$ step by step as follows:
Let $\CalQ_N^{(i)}[F] = ( \mu^{(i)} \circ \cdots \circ \mu^{(1)} ) \CalQ_N[F]$ for $i = 1, 2, 3$.
The first sequence of mutations $\m^{(1)}$ from follow the arrow rule transport one of the handle in the quiver $\CalQ_N[F]$ from the top to the bottom. 
The second sequence of mutations changes the orientation of all the triangles in the quiver $\CalQ_N[F]^{(1)}$. 
If we rotate the quiver $\CalQ_N[F]^{(2)}$ by 180 degrees, $\CalQ_N[F]^{(2)}$ becomes a quiver that is similar to the original $\CalQ_N[F]$ with just minor differences: both handles are now on the top of the quiver, and the last flavor node (if it exists) is also moved to the top of the quiver. 

The second \emph{follow the arrow rule} will transport the other handle to the top of the quiver $\CalQ_N[F]^{(2)}$. 
Finally, the last sequence of mutations will change the orientation of all the triangles in the quiver $\CalQ_N[F]^{(3)}$ again, effectively canceling the change made by $\m^{(2)}$. We recover the original quiver up to a permutation of vertices. 
\end{proof}

\begin{Remark}
The $\bQ$-operator defined at \cref{def:Baxter-Q} corresponds to the \emph{top} Baxter $\bQ$-operator in \cite{schrader2018b}, in the sense that it reproduces the case $F=\varnothing$. Similarly, we can define the \emph{bottom} Baxter $\bQ$-operator whose inverse is the second follow the arrow mutation sequence $\m^{(3)}$ in \cref{def:second-follow-the-arrow}. We will focus on the top Baxter $\bQ$-operator in this paper. 
\end{Remark}

\subsection{Integral operators}\label{sec:int_op}

We now explain the action of the Baxter $\bQ$-operator defined by the sequence of mutations on the functional space.

Let $\CalF$ be the space of entire functions $f$ such that
\begin{align*}
    \int_\BR |f(x+iy)|^2 dx < \infty\quad \forall ~ y \in \BR .
\end{align*}
The space $\CalF$ is dense in $L^2(\BR)$ and clearly contains the subspace
\[
    \{e^{-\a x^2+\b x}g(x)~|~\a\in \BR_{>0},~\b\in \BC,~ g(x)\in \BC[x]\} .
\]
We write $\CalF_N = \CalF^{\otimes N}$ the space of entire functions with $N$ variables, which is the representation of the Heisenberg $\ast$-algebra on $L^2(\mathbb{R}^N)$. 

We consider the action of the operator $\varphi(p+u)$, $u \in \mathbb{R}$.
Using \eqref{eq:qdf-fourier}, it is expressed as an integral operator acting on $f \in \CalF$:
\begin{align}\nonumber
\begin{split}
    \varphi(p+u)f(x) & = \zeta^{-1}\int_{\BR+i0} \frac{e^{2\pi i t(p+u-c_b)}}{\varphi(t-c_b)} f(x)dt \\
    & = \zeta^{-1}\int_{\BR+i0} \frac{e^{2\pi i t(u-c_b)}}{\varphi(t-c_b)} f(x+t) dt .
\end{split}
\end{align}
See \cref{sec:functions} for the definition of $\varphi$, $c_b$, and $\zeta$.
Shifting $t \mapsto t-x$ gives 
\begin{align}\nonumber
    \varphi(p+u)f(x) = \zeta^{-1}e^{2\pi i x(c_b-u)}\int_{\BR+i0} \frac{e^{2\pi it (u+c_b)}}{\varphi(t-x-c_b)} f(t) dt .
\end{align}

\begin{Lemma}\label{lem:T-as-operator}
    Let $f\in \CalF_N$. For $\{T_k\}_{k=1}^N$ defined in \cref{prop:Baxter-Q}, we have 
    \begin{align*}
        T_k(u) f(\bx) = \varphi(x_{k}-x_{k-1}+u-m_k)^{\mathbf{1}_F(k)} \zeta^{-1} \int_{\mathbb{R}+i0}  \frac{e^{2\pi i(y_k-x_k)(u-c_b)}\varphi(y_k-x_{k-1})}{\varphi(y_k-x_k-c_b)\varphi(x_k-x_{k-1})} f(\bx)|_{x_k\to y_k} dy_k
    \end{align*}
    where
    \[
        f(\bx)|_{x_k\to y_k} = f(x_1,\dots,x_{k-1},y_k,x_{k+1},\dots,x_N).
    \]
\end{Lemma}
\begin{proof}
    See \cref{proof:lem:T-as-operator}.
\end{proof}

\begin{Proposition}\label{cor:Baxter-integral}
    The Baxter $\bQ$-operator of $\CalQ_N[F]$ acts on $\CalF_N$ by
    \begin{align}\nonumber
    \begin{split}
        \bQ_N(u) f(\bx) = \zeta^{-N} & \int_{(\BR+i0)^N}  e^{2\pi i (\underline{\bx}-\underline{\by})(c_b-u)} \prod_{k=1}^N \frac{1}{\varphi(y_k-x_k-c_b)} \\
        & \quad \times \prod_{k=1}^{N-1} \frac{\varphi(x_{k+1}-y_k+u-m_{k+1})^{\mathbf{1}_F(k)}\varphi(y_{k+1}-y_k)}{\varphi(x_{k+1}-y_k)}f(\by) d\by. 
    \end{split}
    \end{align}
    where $\bx\in \BR^N$. $\underline{\bx}=x_1+\cdots+x_N$, $\underline{\by}=y_1+\cdots+y_{N}$.
\end{Proposition}
\begin{proof}
    It follows from \cref{lem:T-as-operator}. 
\end{proof}

\begin{Definition}{}\label{def:Q-function}
    Let $\d_k = x_{k+1}-x_{k}$ for $k=1,\dots,N-1$.% 
    \footnote{This corresponds to the simple root of type $A_N$.} We define the $Q$-function as the Fourier transform of the Baxter $\bQ$-operator acting on test function 1 by
    \begin{align*}
        Q_{\ba}(u) = \zeta^{1-N}\int_{(\BR+i0)^{N-1}} \bQ_N(u) 1 \prod_{k=1}^{N-1} e^{2\pi i\d_k a_k} \frac{d\d_k}{2\pi i} .
    \end{align*}
    with ${\ba}=(a_1,\dots,a_{N-1},a_N=0) \in \mathbb{R}^N$.
\end{Definition}
We remark that even though $1 \notin \CalF_N$, the definition of the $Q$-function by Fourier transform in \cref{def:Q-function} yields finite result in the integration. 
\begin{Proposition}\label{prop:Q-function-quiver}
    For $\CalQ_{N}[F]$ the $Q$-function takes the form
    \begin{align*}
    \begin{split}
        Q_{\ba}(u) = & ~\prod_{k=1}^{N} e^{-2\pi iua_k} \varphi(u+a_k) \prod_{k=2}^{N} e^{-2\pi i a_{k-1}a_k}  \varphi(a_{k-1}+c_b) \left[\frac{\varphi(-a_k-m_k)}{\varphi(a_{k-1}-a_k-m_k)}\right]^{\mathbf{1}_F(k)}.
    \end{split}
    \end{align*} 
\end{Proposition}

\begin{proof}
    See \cref{proof:prop:Q-function-quiver}. 
\end{proof}

\section{Quantum cluster integrable systems}\label{sec:Integrable}

We now demonstrate how the quiver $\CalQ_{N}$ defines a quantum cluster integrable system through the Baxter $\bQ$-operator defined in \cref{def:Baxter-Q}. 

\begin{Lemma}\label{lem:TT}
    The following identity holds for any $u,v \in \BR$:
    \begin{align*}
        T_k(u)T_k(v) = T_k(v)T_k(u) , \quad k = 2, \ldots, N .
    \end{align*}
\end{Lemma}
\begin{proof}
    The statement is a direct consequence of the pentagon identity \eqref{eq:qdf-pentagon} together with the definition of $T_k$ given in \cref{prop:Baxter-Q}. 
    See \cref{proof:lem:TT} for details. 
\end{proof}

For a quiver $\CalQ_N$, we define a partial $\bQ$-operator whose inverse is a subpart of $\m^{(1)}$
\begin{align*}
    \bQ_k(u)^{-1} := \m_{3k-3} \cdots \m_1 \m_0~,~ k=2,\dots,N-1, 
\end{align*} 
which can be written in terms of $\{T_j\}_{j=2,\ldots,k}$,
\begin{align} \label{eq:partial_Baxter_Q}
    \bQ_k(u) = \bQ_1(u)T_2(u) \cdots T_k(u).
\end{align}
We remark that the partial operator $\bQ_k(u)$ of $\CalQ_{N}$ is the full Baxter $\bQ$-operator of a shorter quiver $\CalQ_{k}$.

\begin{Proposition}\label{lem:QQ}
    The $\bQ$-operator forms commutative families. For any $u,v\in \BR$ we have
    \begin{align*}
        \bQ_k(u) \bQ_k(v) = \bQ_k(v) \bQ_k(u).
    \end{align*}
\end{Proposition}
\begin{proof}
See \cref{proof:lem:QQ}.
\end{proof}

\begin{Proposition}\label{prop:t-recursive}
Given a partial $\bQ$-operator, the partial transfer matrix by the partial $\bQ$-operators is an operator-valued polynomial in $e^{2\pi bu}$,
\begin{align*}
    \mathring{\bt}_k(u) := \bQ_k(u-ib/2)\bQ_k(u+ib/2)^{-1} = \sum_{j=0}^k H_{k,j} e^{2\pi bju} , \quad k = 1, \ldots, N .
\end{align*}
Moreover, for any $u,v \in \mathbb{R}$, we have
\begin{align*}
\mathring{\bt}_k(u)\mathring{\bt}_k(v) = \mathring{\bt}_k(v) \mathring{\bt}_k(u) .
\end{align*}
\end{Proposition}
\begin{proof}
Using properties of the quantum dilogarithm function $\varphi(u)$ \eqref{eq:qdf-gamma-like} and \eqref{eq:qdf-passing}, we obtain the recursive relation:
\begin{align}\label{eq:t-recursive-quiver}
\begin{split}
    \mathring{\bt}_k(u) =
    & \left( 1 + e^{2\pi b(p_k+u)} \right) \mathring{\bt}_{k-1}(u) \\
    & + \left( e^{2\pi b(p_k+x_k+u)} + e^{2\pi b(x_k+u-m_k)} \right)\left( \sum_{n=1}^{k-1} e^{\sum\limits_{j=n+1}^{k-1} 2\pi b(u-m_j)} e^{-2\pi bx_{n}} \mathring{\bt}_{n-1}(u)  \right) 
\end{split}
\end{align}
for $k=2,\dots,N$ and $\mathring{\bt}_0(u) = 1$, from which we conclude the polynomiality.
For quivers $\CalQ_N[F]$ with $F\subset\{2,\dots,N\}$, we take $m_k \to \infty $ if $k\notin F$. 
The commutativity immediately follows from \cref{lem:QQ}.
\end{proof}

\begin{Theorem}\label{thm:integrability}
    A quiver $\CalQ_{N}[F]$ and the set of flavor parameters $\bm$ define a cluster integrable system through the (partial) Baxter $\bQ$-operator~\eqref{eq:partial_Baxter_Q}, i.e., the operators appearing in the transfer matrix, that we call the Hamiltonians, form a commuting set of operators
    \begin{align*}
        [H_{j,k},H_{j,l}] = 0 , \quad j = 1, \ldots, N, \ k,l=0,\dots,j.
    \end{align*}
\end{Theorem}

Here we list a few Hamitonians 
\begin{align}\label{def:fT-H-q}
\begin{split}
    & H_{k,0} = 1, \qquad H_{k,k} = e^{2\pi b\sum_{j=1}^k p_j},~ \\
    & H_{k,1} = \sum_{j=1}^k e^{2\pi bp_j} + \sum_{j=2}^{k} \left(  e^{2\pi b(p_j+x_{j}-x_{j-1})}  + {\bf 1}_F(j) e^{2\pi b(x_{j}-x_{j-1}-m_j)} \right).
\end{split}
\end{align}
We call $H_{k,1}$ the Hamiltonian of $q$-deformed open flavored Toda chain. In particular, \eqref{def:fT-H-q} is identified with \eqref{def:fToda-H-q-open} by:
\begin{itemize}
    \item Shift $p_j \mapsto p_j+\frac{i}{2b}$, $x_{j}-x_{j-1}+\frac{ib}{2}$, $m_{j} \mapsto m_j + \frac{ib}{2}$.
    \item Rename $2\pi b p_j \mapsto p_j$, $2\pi bx_j \mapsto x_j$, $2\pi bm_j \mapsto m_j$, $2\pi bu \mapsto -u$. 
    \item The Planck constant $\hbar = 2\pi b^2$.
\end{itemize}
This integrable system has been known to exist \cite{Gorsky:1996hs} but never explicitly written.

\subsection{Sklyanin's Lax formalism} 

We have shown that a quiver $\CalQ_N[F]$ defines a quantum integrable system through the Baxter $\bQ$-operators (\cref{def:Baxter-Q}). 
Let us examine the properties of these quantum cluster integrable systems in more details.

The quantum integrability for spin chains is characterized by the $R$-matrix $R_{\ta_1,\ta_2}(u) \in \operatorname{End} (\EV_{\ta_1} \otimes \EV_{\ta_2})$ which satisfies the Yang-Baxter equation \cite{baxter1985exactly,jimbo1989introduction}
\begin{align}\label{eq:YBE}
    R_{\ta_1,\ta_2}(u-w) R_{\ta_1,\ta_3}(u) R_{\ta_2,\ta_3}(w) = R_{\ta_2,\ta_3}(w) R_{\ta_1,\ta_3}(u) R_{\ta_1,\ta_2}(u-w) \in \operatorname{End} (\EV_{\ta_1}\otimes \EV_{\ta_2}\otimes \EV_{\ta_3})
\end{align}
The Lax operator of rank two is a special case of the $R$-matrix with
\begin{align*}
    \EV_{\ta_1} = \EV_{\ta_2} = \EV_{\rm aux} = \BC^2~,~ \EV_{\ta_3} = \EH_k
\end{align*}
The local Hilbert space of the $k$-th particle is denoted by $\EH_k$. 
We apply Sklyanin's Lax matrices to this one-dimensional lattice model~\cite{sklyanin1995separation}, 
\begin{align*}
    L_k(u) = \begin{pmatrix}
        A_k & B_k \\ C_k & D_k
    \end{pmatrix} \in \operatorname{End}(\EV_\text{aux} \otimes \EH_k) ~,~ k=1,\dots,N ,
\end{align*}
where $A_k,B_k,C_k,D_k \in \operatorname{End}(\EH_k)$. 
The $R$-matrix is given by 
\begin{align*}
    R(u) = \begin{pmatrix}
        \sinh\pi b(u+b) & & & \\
        & \sinh \pi bu & \sinh \pi b^2 & \\
        & \sinh \pi b^2 & \sinh \pi bu & \\
        & & & \sinh \pi b(u+b)
    \end{pmatrix} \in \operatorname{End}(\EV_\text{aux}\otimes \EV_\text{aux}) , \quad b \in \mathbb{R}_{>0} .
\end{align*}
The monodromy matrix is a concatenation of Lax matrices: 
\begin{align*}
     \bT_N(u) = K L_N(u) \cdots L_1(u) \in \operatorname{End} (\EV_\text{aux}\otimes \EH),
\end{align*}
where $\EH=\EH_1 \otimes \cdots \otimes \EH_N$ is the Hilbert space of the whole lattice. 
$K \in \operatorname{End}(\EV_\text{aux})$ is a twist matrix that imposes a twisted periodic boundary condition on the lattice. 
The generating function of the conserving Hamiltonians, known as the transfer matrix, is given by the trace of the monodromy matrix with respect to $\CalV_\text{aux}$,
\begin{align*}
    \bt^{\rm Lax}_N(u) = \Tr_{\EV_\text{aux}} \bT_N(u) .
\end{align*}

An open chain is obtained by choosing a proper twist matrix: 
\begin{align*}
    \mathring{K} = \begin{pmatrix}
        1 & 0 \\ 0 & 0
    \end{pmatrix}.
\end{align*}
We define a partial monodromy matrix and transfer matrix by
\begin{align*}
\mathring{\bT}_k(u)
& = \mathring{K} L_k(u) \cdots L_1(u) ~, \quad \mathring{\bt}^{\rm Lax}_k(u) = \Tr_{\CalV_{\rm aux}} \mathring{\bT}_k(u) 
\end{align*}
with any $1 \le k \le N$. 
It is obvious that the $k$-th partial transfer matrix of an open lattice is the transfer matrix of the open lattice with the first $k$ sites. 
The partial transfer matrix satisfies the recursive relations:
\begin{align}\label{eq:t-recursive}
\begin{split}
    \mathring{\bt}^\text{Lax}_k(u)
    & = A_k \mathring{\bt}^\text{Lax}_{k-1}(u) + B_kC_{k-1} \mathring{\bt}^\text{Lax}_{k-2}(u) + B_kD_{k-1}C_{k-2} \mathring{\bt}_{k-3}^\text{Lax}(u)  \\
    & \qquad + \cdots + B_k D_{k-1}\cdots D_3 C_2 \mathring{\bt}^\text{Lax}_{1}(u)
    +B_k D_{k-1} \cdots D_3D_2C_1
\end{split}
\end{align}
for $k=2,\dots,N$.

Comparing the recursive relation \eqref{eq:t-recursive} of the transfer matrix from the Lax matrices with that obtained from the quiver mutation in \eqref{eq:t-recursive-quiver}, we have the following theorem.
\begin{Theorem}\label{thm:Lax_matrices}
    The integrable system defined by the quiver $\CalQ_{N}[F]$ is an open chain system whose Lax matrices are given by 
    \begin{align*}
    L_k(u) = \begin{pmatrix}
        1 + e^{2\pi b(p_k+u)} & e^{2\pi b(p_k+x_k+u)} + \mathbf{1}_F(k) e^{2\pi b(x_k+u-m_{k})} \\
        e^{-2\pi b x_k} & \mathbf{1}_F(k) e^{2\pi b(u-m_{k})}
    \end{pmatrix}, \quad k = 1, \ldots, N ,
    \end{align*}
    where we denote by $\mathbf{1}_F(k)$ the indication function~\eqref{eq:true-or-false}. The Lax matrices satisfy the Yang-Baxter equation \eqref{eq:YBE}. 
\end{Theorem}
\begin{proof}
A direct computation shows that the Lax matrix satisfies the Yang-Baxter relation. 
\end{proof}

We note here that by shifting $p_k \mapsto p_k - m_k$, the flavor parameters $m_k$'s act as inhomogeneity parameters of the integrable chain.

We remark that when $F=\varnothing$, $L_k(u)$ is the Lax operator for the open $q$-deformed Toda chain. 
The cluster integrable system for generic $F$ is an open chain {interpolating} Heisenberg XXZ spin chain and $q$-deformed Toda chain. This is similar but not equivalent to the $Y^{N,|F|}$ dimer model~\cite{Lee:2023wbf}.

\subsection{Moving flavors across the open chain}

Consider a quiver $\CalQ_N[F]$. Suppose $k+1\in F$ and $k\notin F$. 
    The following mutation
    \[
        \m_{3k} \m_{3k-2}
    \]
    maps $\CalQ_N[F]$ to a new quiver $\CalQ_N[F']$ with a different set of positions $F'=(F \cup \{k\}) \backslash \{k+1\} $. The mutations effectively shift the flavors by one site on the open chain. 

\begin{Theorem}\label{thm:iso}
    Let $(\CalQ_N[F],\bm)$ and $(\CalQ_N[F'],\bm')$ be two quivers with associated sets of flavor parameters. If there exists a bijective map $c:F \to F'$ such that
    \[
        m_k = m'_{c(k)}~,~ \forall k\in F,
    \]
    then the flavored Toda chain defined from $(\CalQ_N[F],\bm)$ and $(\CalQ_N[F'],\bm')$ are canonically equivalent, i.e., the Hamiltonians $H_{N,k}(\bx,\bp)$ (resp. $H_{N,k}'(\bx',\bp')$) defined by the quiver $\CalQ_N[F]$ (resp. $\CalQ_N[F]'$)
    are identical up to a canonical transform of the canonical coordinates $(\bx,\bp)$ to $(\bx',\bp')$. 
\end{Theorem}
\begin{proof}
    The mutation $\m_{3k}\m_{3k-2}$ maps the cluster variables in the initial quiver $\CalQ_N[F]$ to those of $\CalQ_N[F']$:
    \[
    \mu_{3k-2} : (e^{2\pi bv_{3k-3}},e^{2\pi bv_{3k-2}},e^{2\pi bv_{3k-1}},e^{2\pi bv_{3k}},e^{2\pi bv_{3k+1}}) \mapsto (e^{2\pi bv_{3k-3}'}, e^{2\pi bv_{3k}'}, e^{2\pi bv_{3k-2}'}, e^{2\pi bv_{3k-4}'}, e^{2\pi bv_{3k+1}'}) .
    \]
    We write the mutated cluster variables in terms of canonical coordinates by 
    \begin{align*} \nonumber
        & v_{3k-3}' = x_{k}'-x_{k-1}'+p_{k-1}'+m_{k}'~,~v_{3k-2}' = -p_{k}'-m_{k}' ~,~ v_{3k-1}' = x_k - x_{k-1}' ~,~ \\
        & v_{3k}' = x_{k}'-x_{k+1}' ~,~ v_{3k+1}' = x_{k+1}'-x_{k+2}'+p_{k+2}'-p'_{k+1}~.
    \end{align*}
    The mutation can be realized by the following canonical transformation:
    \begin{align*}
        & e^{2\pi bp_{k}'} = \left( 1 + q e^{2\pi b (x_{k+1}-x_{k}-p_{k}-m_{k+1}) } \right) e^{2\pi bp_{k}}~, \quad
        e^{2\pi bp_{k+1}'} = e^{2\pi bp_{k+1}} \left( 1 + q e^{2\pi b (x_{k+1}-x_{k}-p_{k}-m_{k+1}) } \right)^{-1} ~,\\
        & e^{2\pi bx_{k}'} = e^{2\pi bx_{k}} \left( 1 + q e^{2\pi b (x_{k+1}-x_{k}-p_{k}-m_{k+1}) } \right)^{-1} ~, \quad
        e^{2\pi bx_{k+1}'} = e^{2\pi bx_{k+1}}~.
    \end{align*}
    The second mutation $\m_{3k}$ further maps 
    \[
        e^{2\pi bx_k'} \mapsto e^{2\pi bx_k'} + e^{2\pi b(x_k'-p_k'-m_{k}')}.
    \]
    The flavor parameters are identified by $m_k'=m_{k+1} $.
    The product of the Lax matrices at sites $k$ and $k+1$ before and after the mutation is related by
    \begin{align}\nonumber
    \begin{split}
        & \cdots L_{k+1}'(u) L_{k}'(u) \cdots \\
        & = \cdots \begin{pmatrix}
        1 + e^{2\pi b(p_{k+1}'+u)} & e^{2\pi b(p_{k+1}'+x_{k+1}'+u)} \\ e^{-2\pi b x_{k+1}'} & 0
    \end{pmatrix} \begin{pmatrix}
        1 + e^{2\pi b(p_{k}'+u)} & e^{2\pi b(p_{k}'+x_{k}'+u)} + e^{2\pi b(x_k'+u-m_{k}')} \\ 
        e^{-2\pi bx_{k}'} & e^{2\pi b(u-m_{k}')}
    \end{pmatrix}  \cdots \\
    %%%%%
    & = \cdots
    % \begin{pmatrix}
    %      & \\ 
    %      & 
    % \end{pmatrix}
    \\
    %%%%%
    & = \cdots  \begin{pmatrix}
        1 + e^{2\pi b(p_{k+1}+u)} & e^{2\pi b(p_{k+1}+x_{k+1}+u)} + e^{2\pi b(x_{k+1}+u-m_{k+1})} \\ 
        e^{-2\pi bx_{k+1}} & e^{2\pi b(u-m_{k+1})}
    \end{pmatrix} \begin{pmatrix}
        1 + e^{2\pi b(p_{k}+u)} & e^{2\pi b(p_{k}+x_{k}+u)} \\ e^{-2\pi b x_{k}} & 0
    \end{pmatrix} \cdots  \\
    & = \cdots L_{k+1}(u) L_{k}(u) \cdots .
    \end{split}
    \end{align}
    Therefore the mutation is local. All other Lax operators are identical through the mutation $L_n'(u) = L_n(u)$ for $n\neq k,k+1$. 
    Thus 
    \begin{align}\nonumber
    \begin{split}
        \mathring{\bT}_N'(u) 
        & = \begin{pmatrix}
            1 & 0 \\ 0 & 0
        \end{pmatrix} L_{N}'(u) \cdots L_{k+2}'(u)L_{k+1}'(u)L_k'(u)L_{k-1}'(u) \cdots L_1'(u) \\
        & = \begin{pmatrix}
            1 & 0 \\ 0 & 0
        \end{pmatrix} L_{N}(u) \cdots L_{k+2}(u)L_{k+1}(u)L_k(u)L_{k-1}(u) \cdots L_1(u) \\
        & = \mathring{\bT}_N(u).
    \end{split}
    \end{align}
    Hence for the transfer matrix
    \begin{align}\nonumber
        \sum_{k=0}^N H_{N,k}'(\bx',\bp') e^{2\pi bku}= \mathring{\bt}_N'(u) = \Tr \mathring{\bT}_N'(u) = \Tr \mathring{\bT}_N(u)  = \mathring{\bt}_N(u) = \sum_{k=0}^N H_{N,k}(\bx,\bp) e^{2\pi bku}
    \end{align}
    holds for any value of $u\in \BR$. For the Hamiltonians of the two quivers are equivalent up to a canonical transformation of the coordinates:
    \begin{align}\nonumber
        H_{N,k}'(\bx',\bp') = H_{N,k}(\bx,\bp)~,~ k=0,\dots,N.
    \end{align}
    This concludes the proof.
\end{proof}

This equivalence is also observed in the $Y^{N,k}$ quantum cluster integrable model \cite{Lee:2023wbf} (also called $Y^{p,q}$ model in other literature), where multiple different graphs can be constructed for the same $Y^{N,k}$ model. Those graphs are related by cluster mutations, and hence the quantum cluster integrable systems are canonically equivalent.

\section{Examples}\label{sec:Example}

We demonstrate our construction with some examples. 
We use the following notation,
\begin{align}\label{eq:multiplicative_notation}
\begin{split}
    & \a_j = e^{2\pi bp_j}=P_j, \quad j=1,\ldots,N, \\
    & \b_j = e^{2\pi b (p_j+x_j-x_{j-1})} =q P_jX_jX_{j-1}^{-1}, \quad  
    \g_j = e^{2\pi b(x_j-x_{j-1}-m_j)}=X_jX_{j-1}^{-1}M_j^{-1}, \quad j=2,\ldots,N.
\end{split}
\end{align}
We have $\gamma_j \to 0$ in the flavor decoupling limit $m_j \to \infty$.

\subsection{$N=3$}

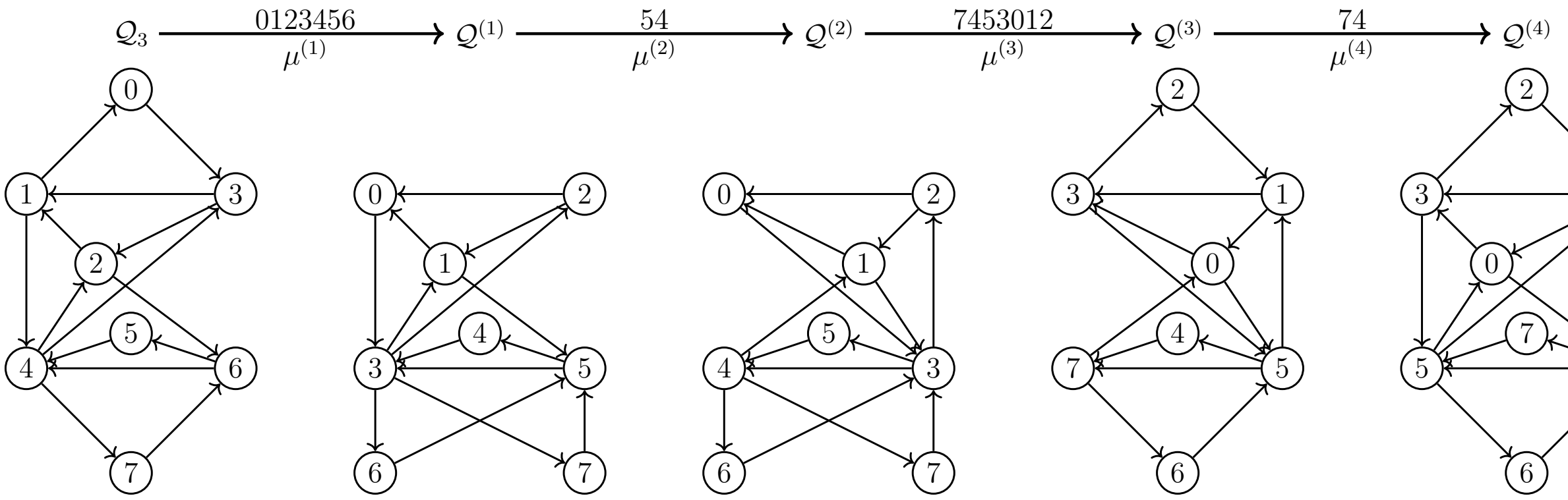
\begin{figure}[h!] 
% \if0
\centering
\scalebox{.7}{
\begin{tikzpicture}[square/.style={regular polygon,regular polygon sides=4},decoration={markings, 
    mark= at position 0.5 with {\arrow[scale=1.3]{stealth}}}
 ]

    %%% Quiver 1 %%%%%%
    
    \node at (-2.5,-1.5) [style={circle,draw},thick,inner sep=6pt] (r0) {};
    \node at (-4,-3) [style={circle,draw},thick,inner sep=6pt] (r1) {};
    \node at (-3,-4) [style={circle,draw},thick,inner sep=6pt] (r3) {};
    \node at (-1,-3) [style={circle,draw},thick,inner sep=6pt] (r2) {};
    \node at (-4,-5.5) [style={circle,draw},thick,inner sep=6pt] (r4) {};
    \node at (-1,-5.5) [style={circle,draw},thick,inner sep=6pt] (r5) {};
    \node at (-2.5,-5) [style={circle,draw},thick,inner sep=6pt] (r6) {};
    \node at (-2.5,-7) [style={circle,draw},thick,inner sep=6pt] (r7) {};

    \node at (-2.5,-.7) (Q1){$\CalQ_3$};
    \node at (-2.5,-1.5) {0};
    \node at (-4,-3) {1};
    \node at (-1,-3) {3};
    \node at (-3,-4) {2};
    \node at (-4,-5.5) {4};
    \node at (-1,-5.5) {6};
    \node at (-2.5,-5) {5};
    \node at (-2.5,-7) {7};

    %%% Arrows %%%

    \draw[thick,->] (r1) -- (r0);
    \draw[thick,->] (r0) -- (r2);
    \draw[thick,->] (r2) -- (r1);
    \draw[thick,->] (r1) -- (r4);
    \draw[thick,->] (r4) -- (r2); 
    \draw[thick,->] (r2) -- (r3);
    \draw[thick,->] (r3) -- (r1);
    \draw[thick,->] (r4) -- (r3);
    \draw[thick,->] (r3) -- (r5);
    \draw[thick,->] (r5) -- (r4); 
    \draw[thick,->] (r4) -- (r7);
    \draw[thick,->] (r6) -- (r4);
    \draw[thick,->] (r5) -- (r6);
    \draw[thick,->] (r7) -- (r5);

    %%% QUIVER 2 %%%
    
    % \node at (-2.5,-1.5) [style={circle,draw},thick,inner sep=6pt] (r0) {};
    \node at (1,-3) [style={circle,draw},thick,inner sep=6pt] (q1) {};
    \node at (2,-4) [style={circle,draw},thick,inner sep=6pt] (q3) {};
    \node at (4,-3) [style={circle,draw},thick,inner sep=6pt] (q2) {};
    \node at (1,-5.5) [style={circle,draw},thick,inner sep=6pt] (q4) {};
    \node at (4,-5.5) [style={circle,draw},thick,inner sep=6pt] (q5) {};
    \node at (2.5,-5) [style={circle,draw},thick,inner sep=6pt] (q6) {};
    \node at (1,-7) [style={circle,draw},thick,inner sep=6pt] (q7) {};
    \node at (4,-7) [style={circle,draw},thick,inner sep=6pt] (q8) {};

    \node at (2.5,-.7) (Q2){$\mathcal{Q}^{(1)}$};
    % \node at (2.5,-1.5) {0};
    \node at (1,-3) {0};
    \node at (4,-3) {2};
    \node at (2,-4) {1};
    \node at (1,-5.5) {3};
    \node at (4,-5.5) {5};
    \node at (2.5,-5) {4};
    \node at (1,-7) {6};
    \node at (4,-7) {7};

    %%% Arrows %%%

    % \draw[thick,->] (r1) -- (r0);
    % \draw[thick,->] (r0) -- (r2);
    \draw[thick,->] (q2) -- (q1);
    \draw[thick,->] (q1) -- (q4);
    \draw[thick,->] (q4) -- (q2); 
    \draw[thick,->] (q2) -- (q3);
    \draw[thick,->] (q3) -- (q1);
    \draw[thick,->] (q4) -- (q3);
    \draw[thick,->] (q3) -- (q5);
    \draw[thick,->] (q5) -- (q4); 
    \draw[thick,->] (q4) -- (q7);
    \draw[thick,->] (q6) -- (q4);
    \draw[thick,->] (q5) -- (q6);
    \draw[thick,->] (q7) -- (q5);

    \draw[thick,->] (q4) -- (q8);
    \draw[thick,->] (q8) -- (q5);

    %%% QUIVER 3 %%%
    
    % \node at (-2.5,-1.5) [style={circle,draw},thick,inner sep=6pt] (r0) {};
    \node at (6,-3) [style={circle,draw},thick,inner sep=6pt] (p1) {};
    \node at (8,-4) [style={circle,draw},thick,inner sep=6pt] (p3) {};
    \node at (9,-3) [style={circle,draw},thick,inner sep=6pt] (p2) {};
    \node at (6,-5.5) [style={circle,draw},thick,inner sep=6pt] (p4) {};
    \node at (9,-5.5) [style={circle,draw},thick,inner sep=6pt] (p5) {};
    \node at (7.5,-5) [style={circle,draw},thick,inner sep=6pt] (p6) {};
    \node at (6,-7) [style={circle,draw},thick,inner sep=6pt] (p7) {};
    \node at (9,-7) [style={circle,draw},thick,inner sep=6pt] (p8) {};

    \node at (7.5,-.7) (Q3){$\mathcal{Q}^{(2)}$};
    % \node at (2.5,-1.5) {0};
    \node at (6,-3) {0};
    \node at (9,-3) {2};
    \node at (8,-4) {1};
    \node at (6,-5.5) {4};
    \node at (9,-5.5) {3};
    \node at (7.5,-5) {5};
    \node at (6,-7) {6};
    \node at (9,-7) {7};

    %%% Arrows %%%

    % \draw[thick,->] (r1) -- (r0);
    % \draw[thick,->] (r0) -- (r2);
    \draw[thick,->] (p1) -- (p5);
    \draw[thick,->] (p2) -- (p1); 
    \draw[thick,->] (p2) -- (p3);
    \draw[thick,->] (p3) -- (p1);
    \draw[thick,->] (p4) -- (p3);
    \draw[thick,->] (p3) -- (p5);
    \draw[thick,->] (p5) -- (p4);
    \draw[thick,->] (p5) -- (p2);
    \draw[thick,->] (p4) -- (p7);
    \draw[thick,->] (p6) -- (p4);
    \draw[thick,->] (p5) -- (p6);
    \draw[thick,->] (p7) -- (p5);

    \draw[thick,->] (p4) -- (p8);
    \draw[thick,->] (p8) -- (p5);

    %%% QUIVER 4 %%%
    
    \node at (12.5,-1.5) [style={circle,draw},thick,inner sep=6pt] (t0) {};
    \node at (11,-3) [style={circle,draw},thick,inner sep=6pt] (t1) {};
    \node at (13,-4) [style={circle,draw},thick,inner sep=6pt] (t3) {};
    \node at (14,-3) [style={circle,draw},thick,inner sep=6pt] (t2) {};
    \node at (11,-5.5) [style={circle,draw},thick,inner sep=6pt] (t4) {};
    \node at (14,-5.5) [style={circle,draw},thick,inner sep=6pt] (t5) {};
    \node at (12.5,-5) [style={circle,draw},thick,inner sep=6pt] (t6) {};
    \node at (12.5,-7) [style={circle,draw},thick,inner sep=6pt] (t7) {};
    % \node at (9,-7) [style={circle,draw},thick,inner sep=6pt] (p8) {};

    \node at (12.5,-.7) (Q4){$\mathcal{Q}^{(3)}$};
    \node at (12.5,-1.5) {2};
    \node at (11,-3) {3};
    \node at (14,-3) {1};
    \node at (13,-4) {0};
    \node at (11,-5.5) {7};
    \node at (14,-5.5) {5};
    \node at (12.5,-5) {4};
    \node at (12.5,-7) {6};
    % \node at (9,-7) {7};

    %%% Arrows %%%

    \draw[thick,->] (t1) -- (t0);
    \draw[thick,->] (t0) -- (t2);
    \draw[thick,->] (t1) -- (t5);
    \draw[thick,->] (t2) -- (t1); 
    \draw[thick,->] (t2) -- (t3);
    \draw[thick,->] (t3) -- (t1);
    \draw[thick,->] (t4) -- (t3);
    \draw[thick,->] (t3) -- (t5);
    \draw[thick,->] (t5) -- (t4);
    \draw[thick,->] (t5) -- (t2);
    \draw[thick,->] (t4) -- (t7);
    \draw[thick,->] (t6) -- (t4);
    \draw[thick,->] (t5) -- (t6);
    \draw[thick,->] (t7) -- (t5);

    % \draw[thick,->] (p4) -- (p8);
    % \draw[thick,->] (p8) -- (p5);

    %%% Quiver 5 %%%%%%
    
    \node at (17.5,-1.5) [style={circle,draw},thick,inner sep=6pt] (s0) {};
    \node at (16,-3) [style={circle,draw},thick,inner sep=6pt] (s1) {};
    \node at (17,-4) [style={circle,draw},thick,inner sep=6pt] (s3) {};
    \node at (19,-3) [style={circle,draw},thick,inner sep=6pt] (s2) {};
    \node at (16,-5.5) [style={circle,draw},thick,inner sep=6pt] (s4) {};
    \node at (19,-5.5) [style={circle,draw},thick,inner sep=6pt] (s5) {};
    \node at (17.5,-5) [style={circle,draw},thick,inner sep=6pt] (s6) {};
    \node at (17.5,-7) [style={circle,draw},thick,inner sep=6pt] (s7) {};

    \node at (17.5,-.7) (Q5){$\mathcal{Q}^{(4)}$};
    \node at (17.5,-1.5) {2};
    \node at (16,-3) {3};
    \node at (19,-3) {1};
    \node at (17,-4) {0};
    \node at (16,-5.5) {5};
    \node at (19,-5.5) {4};
    \node at (17.5,-5) {7};
    \node at (17.5,-7) {6};

    %%% Arrows %%%

    \draw[thick,->] (s1) -- (s0);
    \draw[thick,->] (s0) -- (s2);
    \draw[thick,->] (s2) -- (s1);
    \draw[thick,->] (s1) -- (s4);
    \draw[thick,->] (s4) -- (s2); 
    \draw[thick,->] (s2) -- (s3);
    \draw[thick,->] (s3) -- (s1);
    \draw[thick,->] (s4) -- (s3);
    \draw[thick,->] (s3) -- (s5);
    \draw[thick,->] (s5) -- (s4); 
    \draw[thick,->] (s4) -- (s7);
    \draw[thick,->] (s6) -- (s4);
    \draw[thick,->] (s5) -- (s6);
    \draw[thick,->] (s7) -- (s5);

    %%%%%%%
    \draw[very thick,->] (Q1) -- (Q2);
    \draw[very thick,->] (Q2) -- (Q3);
    \draw[very thick,->] (Q3) -- (Q4);
    \draw[very thick,->] (Q4) -- (Q5);

    \node at (5,-.5) {54};
    \node at (0,-.5) {0123456};
    \node at (10,-.5) {7453012};
    \node at (15,-.5) {74};
    \node at (0,-1) {$\m^{(1)}$};
    \node at (10,-1) {$\m^{(3)}$};
    \node at (5,-1) {$\m^{(2)}$};
    \node at (15,-1) {$\m^{(4)}$};

\end{tikzpicture}
}
% \fi
\caption{Mutational sequence of the $\CalQ_3$ chain.} 
\label{fig:SU4+2F-open-2}
\end{figure}

The quiver $\CalQ_3=\CalQ_3[\{2,3\}]$ and the mutations $\m^{(r)}$, $r=1,2,3,4$, are shown in \cref{fig:SU4+2F-open-2}.

The first \emph{follow the arrow rule} sequence of mutation $\m^{(1)}$ defines the Baxter $\bQ$-operator and its Fourier transform ($Q$-function):
\begin{align*}
\begin{split}
    \bQ_3(u) 
    & = \varphi(p_1+u) \varphi(x_2-x_1+u-m_2) \varphi(x_2-x_1+p_2+u) \varphi(p_2+u) \\
    & \qquad ~ \varphi(x_3-x_2+u-m_3) \varphi(x_3-x_2+p_3+u) \varphi(p_3+u),  \\ 
    %%%%%
    Q_{\ba}(u) & = \int_{\BR+i0} \frac{d\d_1 d\d_2}{(2\pi i)^2} e^{2\pi i (a_1\d_1+a_2\d_2)} \bQ_3(u)1 \\
    & = e^{-2\pi i u(a_1+a_2)} \varphi(u)\varphi(u+a_1) \varphi(u+a_2) \\
    & \quad \times e^{-2\pi i a_1a_2} \varphi(a_1+c_b)\varphi(a_2+c_b) \frac{\varphi(-m_2-a_2)}{\varphi(a_1-a_2-m_2)} \frac{\varphi(-m_3)}{\varphi(a_2-m_3)}. 
\end{split}
\end{align*}
The Lax matrices are given by 
\begin{align*}
\begin{split}
    & L_1(u) = \begin{pmatrix}
        1 + e^{2\pi b(p_1+u)} & e^{2\pi b(p_1+x_1+u)}  \\
        e^{-2\pi b x_1} & 0
    \end{pmatrix}~,~ \\
    & L_k(u) = \begin{pmatrix}
        1 + e^{2\pi b(p_k+u)} & e^{2\pi b(p_k+x_k+u)} + e^{2\pi b(x_k+u-m_{k})} \\
        e^{-2\pi b x_k} & e^{2\pi b(u-m_{k})}
    \end{pmatrix}~, \quad k=2,3.
\end{split}
\end{align*}

To obtain an open chain with less flavors $\CalQ_3[\{2\}]$ (resp. $\CalQ_3[\{3\}]$), we merge vertex 5 (resp. 2) to vertex 4 (resp. 1) in $\CalQ_3[\{2,3\}]$ in \cref{fig:SU4+2F-open-2}. The merging also takes $m_3\to \infty$ (resp. $m_2 \to \infty$) in the $\bQ$-operator and $Q$-function. 
Using the properties of the quantum dilogarithm function 
\[
    \lim_{z \to -\infty} \varphi(z) = 1,
\]
we obtain the Baxter $\bQ$-operator and $Q$-function for $\CalQ_3[\{2\}]$:
\begin{align*}
%\begin{split}
    \bQ_3(u) 
    & = \varphi(p_1+u) \varphi(x_2-x_1+u-m_2) \varphi(x_2-x_1+p_2+u) \varphi(p_2+u) \\
    & \quad \times \varphi(x_3-x_2+p_3+u) \varphi(p_3+u),  \\ 
    %%%%%
    Q_\ba(u) 
    & = e^{-2\pi i u(a_1+a_2)} \varphi(u)\varphi(u+a_1) \varphi(u+a_2) e^{-2\pi i a_1a_2} \\
    & \quad \times \varphi(a_1+c_b) \varphi(a_2+c_b) \frac{\varphi(-m_2-a_2)}{\varphi(a_1-a_2-m_2)} .
%\end{split}
\end{align*}
As for $\CalQ_3[\{3\}]$, the Baxter $\bQ$-operator and $Q$-function are 
\begin{align*}
%\begin{split}
    \bQ_3(u) 
    & = \varphi(p_1+u) \varphi(x_2-x_1+p_2+u) \varphi(p_2+u) \\
    & \quad \times \varphi(x_3-x_2+u-m_3) \varphi(x_3-x_2+p_3+u) \varphi(p_3+u) , \\ 
    %%%%%
    Q_\ba(u) 
    & = e^{-2\pi i u(a_1+a_2)} \varphi(u)\varphi(u+a_1) \varphi(u+a_2) e^{-2\pi i a_1a_2} \\
    & \quad \times \varphi(a_1+c_b)\varphi(a_2+c_b)  \frac{\varphi(-m_3)}{\varphi(a_2-m_3)} .
%\end{split}
\end{align*}
We have the transfer matrix 
\begin{align*}
    \mathring\bt_3(u) = \bQ_3(u-ib/2) \bQ_3(u+ib/2)^{-1} = 1 + \sum_{k=1}^3 H_{3,k} e^{2\pi bku} ,
\end{align*}
where the Hamiltonians are given by
\begin{subequations}
\begin{align*}
    H_{3,1} = &~ \sum_{j=1}^3 \alpha_j + \sum_{j=2}^3 (\beta_j + \gamma_j) , \\
    %%%%%
    H_{3,2} = &~ \a_1\a_2 + \a_1\a_3 + \a_2\a_3 + \b_2 \a_3 + \b_3\a_1 \\
    & + \g_2\a_3 + \g_3\a_1 + \b_3\g_2 + \g_2\g_3 , \nonumber \\
    %%%%%%
    H_{3,3} = & ~ \a_1\a_2\a_3 .
\end{align*}
\end{subequations}

\subsection{$N=4$}

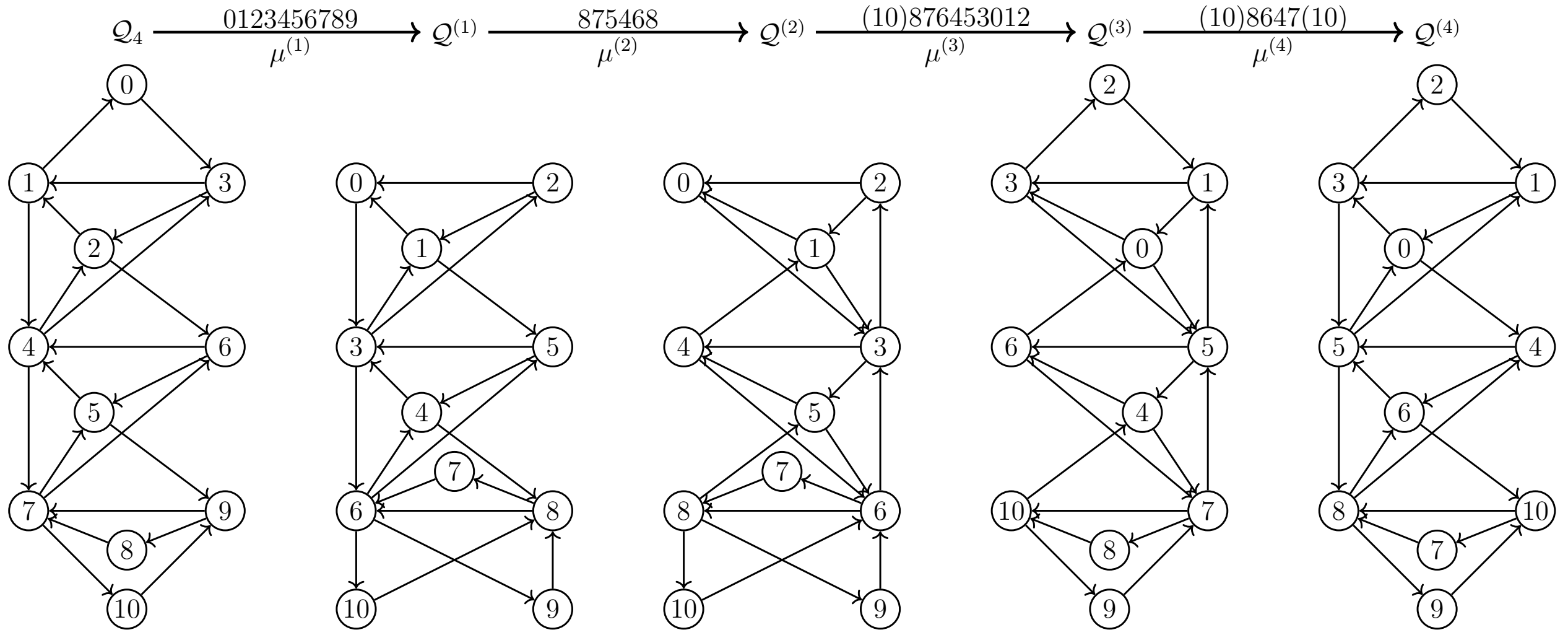
\begin{figure}[h!] 
% \if0
\centering
\scalebox{.7}{
\begin{tikzpicture}[square/.style={regular polygon,regular polygon sides=4},decoration={markings, 
    mark= at position 0.5 with {\arrow[scale=1.3]{stealth}}}
 ]

    %%%%% QUIVER 1 %%%%%

    \node at (-12.5,-1.5) [style={circle,draw},thick,inner sep=6pt] (r0) {};
    \node at (-14,-3) [style={circle,draw},thick,inner sep=6pt] (r1) {};
    \node at (-13,-4) [style={circle,draw},thick,inner sep=6pt] (r3) {};
    \node at (-11,-3) [style={circle,draw},thick,inner sep=6pt] (r2) {};
    \node at (-14,-5.5) [style={circle,draw},thick,inner sep=6pt] (r4) {};
    \node at (-11,-5.5) [style={circle,draw},thick,inner sep=6pt] (r5) {};
    \node at (-13,-6.5) [style={circle,draw},thick,inner sep=6pt] (r6) {};
    \node at (-14,-8) [style={circle,draw},thick,inner sep=6pt] (r7) {};
    \node at (-11,-8) [style={circle,draw},thick,inner sep=6pt] (r8) {};
    \node at (-12.5,-8.6) [style={circle,draw},thick,inner sep=6pt] (r9) {};
    \node at (-12.5,-9.5) [style={circle,draw},thick,inner sep=6pt] (r10) {};

    \node at (-12.5,-.7) (Q1){$\mathcal{Q}_4$};
    \node at (-12.5,-1.5) {0};
    \node at (-14,-3) {1};
    \node at (-11,-3) {3};
    \node at (-13,-4) {2};
    \node at (-14,-5.5) {4};
    \node at (-11,-5.5) {6};
    \node at (-13,-6.5) {5};
    \node at (-14,-8) {7};
    \node at (-11,-8) {9};
    \node at (-12.5,-8.6) {8};
    \node at (-12.5,-9.5) {10};

    %%%%% Arrow %%%%%%

    \draw[thick,->] (r0) -- (r2);
    \draw[thick,->] (r1) -- (r0);
    \draw[thick,->] (r1) -- (r4);
    \draw[thick,->] (r2) -- (r1);
    \draw[thick,->] (r2) -- (r3);
    \draw[thick,->] (r3) -- (r1);
    \draw[thick,->] (r3) -- (r5);
    \draw[thick,->] (r4) -- (r2);
    \draw[thick,->] (r4) -- (r3);
    \draw[thick,->] (r4) -- (r7);
    \draw[thick,->] (r5) -- (r4);
    \draw[thick,->] (r5) -- (r6);
    \draw[thick,->] (r6) -- (r4);
    \draw[thick,->] (r6) -- (r8);
    \draw[thick,->] (r7) -- (r5);
    \draw[thick,->] (r7) -- (r6);
    \draw[thick,->] (r8) -- (r9);
    \draw[thick,->] (r8) -- (r7);
    \draw[thick,->] (r9) -- (r7);

    \draw[thick,->] (r7) -- (r10);
    \draw[thick,->] (r10) -- (r8);

    %%%%% QUIVER 2 %%%%%

    % \node at (-7.5,-1.5) [style={circle,draw},thick,inner sep=6pt] (p0) {};
    \node at (-9,-3) [style={circle,draw},thick,inner sep=6pt] (p1) {};
    \node at (-8,-4) [style={circle,draw},thick,inner sep=6pt] (p3) {};
    \node at (-6,-3) [style={circle,draw},thick,inner sep=6pt] (p2) {};
    \node at (-9,-5.5) [style={circle,draw},thick,inner sep=6pt] (p4) {};
    \node at (-6,-5.5) [style={circle,draw},thick,inner sep=6pt] (p5) {};
    \node at (-8,-6.5) [style={circle,draw},thick,inner sep=6pt] (p6) {};
    \node at (-9,-8) [style={circle,draw},thick,inner sep=6pt] (p7) {};
    \node at (-6,-8) [style={circle,draw},thick,inner sep=6pt] (p8) {};
    \node at (-7.5,-7.4) [style={circle,draw},thick,inner sep=6pt] (p9) {};
    \node at (-6,-9.5) [style={circle,draw},thick,inner sep=6pt] (p10) {};
    \node at (-9,-9.5) [style={circle,draw},thick,inner sep=6pt] (p11) {};

    \node at (-7.5,-.7) (Q2){$\mathcal{Q}^{(1)}$};
    % \node at (-7.5,-1.5) {0};
    \node at (-9,-3) {0};
    \node at (-6,-3) {2};
    \node at (-8,-4) {1};
    \node at (-9,-5.5) {3};
    \node at (-6,-5.5) {5};
    \node at (-8,-6.5) {4};
    \node at (-9,-8) {6};
    \node at (-6,-8) {8};
    \node at (-7.5,-7.4) {7};
    \node at (-6,-9.5) {9};
    \node at (-9,-9.5) {10};

    %%%%% Arrow %%%%%%

    % \draw[thick,->] (p0) -- (p2);
    % \draw[thick,->] (p1) -- (p0);
    \draw[thick,->] (p1) -- (p4);
    \draw[thick,->] (p2) -- (p1);
    \draw[thick,->] (p2) -- (p3);
    \draw[thick,->] (p3) -- (p1);
    \draw[thick,->] (p3) -- (p5);
    \draw[thick,->] (p4) -- (p2);
    \draw[thick,->] (p4) -- (p3);
    \draw[thick,->] (p4) -- (p7);
    \draw[thick,->] (p5) -- (p4);
    \draw[thick,->] (p5) -- (p6);
    \draw[thick,->] (p6) -- (p4);
    \draw[thick,->] (p6) -- (p8);
    \draw[thick,->] (p7) -- (p5);
    \draw[thick,->] (p7) -- (p6);
    \draw[thick,->] (p8) -- (p9);
    \draw[thick,->] (p8) -- (p7);
    \draw[thick,->] (p9) -- (p7);

    \draw[thick,->] (p7) -- (p10);
    \draw[thick,->] (p10) -- (p8);
    \draw[thick,->] (p7) -- (p11);
    \draw[thick,->] (p11) -- (p8);

    %%%%%%
    
    \draw[very thick,->] (Q1) -- (Q2);
    \node at (-10,-.5) {0123456789};
    \node at (-10,-1) {$\m^{(1)}$};

    %%%%%%

    % \draw[very thick,red,->] (-4.7,-9) -- (-4.7,-2);
    % \draw[very thick,red,->] (4.7,-2) -- (4.7,-9);

    %%% QUIVER 3 %%%%%%
    
    % \node at (-2.5,-1.5) [style={circle,draw},thick,inner sep=6pt] (q0) {};
    \node at (-4,-3) [style={circle,draw},thick,inner sep=6pt] (q1) {};
    \node at (-2,-4) [style={circle,draw},thick,inner sep=6pt] (q3) {};
    \node at (-1,-3) [style={circle,draw},thick,inner sep=6pt] (q2) {};
    \node at (-4,-5.5) [style={circle,draw},thick,inner sep=6pt] (q4) {};
    \node at (-1,-5.5) [style={circle,draw},thick,inner sep=6pt] (q5) {};
    \node at (-2,-6.5) [style={circle,draw},thick,inner sep=6pt] (q6) {};
    \node at (-4,-8) [style={circle,draw},thick,inner sep=6pt] (q7) {};
    \node at (-1,-8) [style={circle,draw},thick,inner sep=6pt] (q8) {};
    \node at (-2.5,-7.4) [style={circle,draw},thick,inner sep=6pt] (q9) {};
    \node at (-1,-9.5) [style={circle,draw},thick,inner sep=6pt] (q10) {};
    \node at (-4,-9.5) [style={circle,draw},thick,inner sep=6pt] (q11) {};

    \node at (-2.5,-.7) (Q3){$\mathcal{Q}^{(2)}$};
    % \node at (-2.5,-1.5) {0};
    \node at (-4,-3) {0};
    \node at (-1,-3) {2};
    \node at (-2,-4) {1};
    \node at (-4,-5.5) {4};
    \node at (-1,-5.5) {3};
    \node at (-2,-6.5) {5};
    \node at (-4,-8) {8};
    \node at (-1,-8) {6};
    \node at (-2.5,-7.4) {7};
    \node at (-1,-9.5) {9};
    \node at (-4,-9.5) {10};

    %%% Arrows %%%

    % \draw[thick,->] (q0) -- (q2);
    % \draw[thick,->] (q1) -- (q0);
    \draw[thick,->] (q1) -- (q5);
    \draw[thick,->] (q2) -- (q1);
    \draw[thick,->] (q2) -- (q3);
    \draw[thick,->] (q3) -- (q1);
    \draw[thick,->] (q3) -- (q5);
    \draw[thick,->] (q4) -- (q3);
    \draw[thick,->] (q4) -- (q8);
    \draw[thick,->] (q5) -- (q2);
    \draw[thick,->] (q5) -- (q4);
    \draw[thick,->] (q5) -- (q6);
    \draw[thick,->] (q6) -- (q4);
    \draw[thick,->] (q6) -- (q8);
    \draw[thick,->] (q7) -- (q6);
    \draw[thick,->] (q8) -- (q9);
    \draw[thick,->] (q8) -- (q5);
    \draw[thick,->] (q8) -- (q7);
    \draw[thick,->] (q9) -- (q7);

    \draw[thick,->] (q7) -- (q10);
    \draw[thick,->] (q10) -- (q8);
    \draw[thick,->] (q7) -- (q11);
    \draw[thick,->] (q11) -- (q8);
    
    %%%%%%%%%%

    \draw[very thick,->] (Q2) -- (Q3);
    \node at (-5,-.5) {875468};
    \node at (-5,-1) {$\m^{(2)}$};

    %%%%% QUIVER 4 %%%%%%
    
    \node at (2.5,-1.5) [style={circle,draw},thick,inner sep=6pt] (s0) {};
    \node at (1,-3) [style={circle,draw},thick,inner sep=6pt] (s1) {};
    \node at (3,-4) [style={circle,draw},thick,inner sep=6pt] (s3) {};
    \node at (4,-3) [style={circle,draw},thick,inner sep=6pt] (s2) {};
    \node at (1,-5.5) [style={circle,draw},thick,inner sep=6pt] (s4) {};
    \node at (4,-5.5) [style={circle,draw},thick,inner sep=6pt] (s5) {};
    \node at (3,-6.5) [style={circle,draw},thick,inner sep=6pt] (s6) {};
    \node at (1,-8) [style={circle,draw},thick,inner sep=6pt] (s7) {};
    \node at (4,-8) [style={circle,draw},thick,inner sep=6pt] (s8) {};
    \node at (2.5,-8.6) [style={circle,draw},thick,inner sep=6pt] (s9) {};
    \node at (2.5,-9.5) [style={circle,draw},thick,inner sep=6pt] (s10) {};
    % \node at (-4,-9.5) [style={circle,draw},thick,inner sep=6pt] (q11) {};

    \node at (2.5,-.7) (Q4){$\mathcal{Q}^{(3)}$};
    \node at (2.5,-1.5) {2};
    \node at (1,-3) {3};
    \node at (4,-3) {1};
    \node at (3,-4) {0};
    \node at (1,-5.5) {6};
    \node at (4,-5.5) {5};
    \node at (3,-6.5) {4};
    \node at (1,-8) {10};
    \node at (4,-8) {7};
    \node at (2.5,-8.6) {8};
    \node at (2.5,-9.5) {9};
    % \node at (-4,-9.5) {10};

    %%% Arrows %%%

    \draw[thick,->] (s0) -- (s2);
    \draw[thick,->] (s1) -- (s0);
    \draw[thick,->] (s1) -- (s5);
    \draw[thick,->] (s2) -- (s1);
    \draw[thick,->] (s2) -- (s3);
    \draw[thick,->] (s3) -- (s1);
    \draw[thick,->] (s3) -- (s5);
    \draw[thick,->] (s4) -- (s3);
    \draw[thick,->] (s4) -- (s8);
    \draw[thick,->] (s5) -- (s2);
    \draw[thick,->] (s5) -- (s4);
    \draw[thick,->] (s5) -- (s6);
    \draw[thick,->] (s6) -- (s4);
    \draw[thick,->] (s6) -- (s8);
    \draw[thick,->] (s7) -- (s6);
    \draw[thick,->] (s8) -- (s9);
    \draw[thick,->] (s8) -- (s5);
    \draw[thick,->] (s8) -- (s7);
    \draw[thick,->] (s9) -- (s7);

    \draw[thick,->] (s7) -- (s10);
    \draw[thick,->] (s10) -- (s8);
    % \draw[thick,->] (q7) -- (q11);
    % \draw[thick,->] (q11) -- (q8);
    
    %%%%%%%%%%

    \draw[very thick,->] (Q3) -- (Q4);
    \node at (0,-.5) {(10)876453012};
    \node at (0,-1) {$\m^{(3)}$};

    %%%%% QUIVER 5 %%%%%

    \node at (7.5,-1.5) [style={circle,draw},thick,inner sep=6pt] (t0) {};
    \node at (6,-3) [style={circle,draw},thick,inner sep=6pt] (t1) {};
    \node at (7,-4) [style={circle,draw},thick,inner sep=6pt] (t3) {};
    \node at (9,-3) [style={circle,draw},thick,inner sep=6pt] (t2) {};
    \node at (6,-5.5) [style={circle,draw},thick,inner sep=6pt] (t4) {};
    \node at (9,-5.5) [style={circle,draw},thick,inner sep=6pt] (t5) {};
    \node at (7,-6.5) [style={circle,draw},thick,inner sep=6pt] (t6) {};
    \node at (6,-8) [style={circle,draw},thick,inner sep=6pt] (t7) {};
    \node at (9,-8) [style={circle,draw},thick,inner sep=6pt] (t8) {};
    \node at (7.5,-8.6) [style={circle,draw},thick,inner sep=6pt] (t9) {};
    \node at (7.5,-9.5) [style={circle,draw},thick,inner sep=6pt] (t10) {};

    \node at (7.5,-.7) (Q5){$\mathcal{Q}^{(4)}$};
    \node at (7.5,-1.5) {2};
    \node at (6,-3) {3};
    \node at (9,-3) {1};
    \node at (7,-4) {0};
    \node at (6,-5.5) {5};
    \node at (9,-5.5) {4};
    \node at (7,-6.5) {6};
    \node at (6,-8) {8};
    \node at (9,-8) {10};
    \node at (7.5,-8.6) {7};
    \node at (7.5,-9.5) {9};

    %%%%% Arrow %%%%%%

    \draw[thick,->] (t0) -- (t2);
    \draw[thick,->] (t1) -- (t0);
    \draw[thick,->] (t1) -- (t4);
    \draw[thick,->] (t2) -- (t1);
    \draw[thick,->] (t2) -- (t3);
    \draw[thick,->] (t3) -- (t1);
    \draw[thick,->] (t3) -- (t5);
    \draw[thick,->] (t4) -- (t2);
    \draw[thick,->] (t4) -- (t3);
    \draw[thick,->] (t4) -- (t7);
    \draw[thick,->] (t5) -- (t4);
    \draw[thick,->] (t5) -- (t6);
    \draw[thick,->] (t6) -- (t4);
    \draw[thick,->] (t6) -- (t8);
    \draw[thick,->] (t7) -- (t5);
    \draw[thick,->] (t7) -- (t6);
    \draw[thick,->] (t8) -- (t9);
    \draw[thick,->] (t8) -- (t7);
    \draw[thick,->] (t9) -- (t7);

    \draw[thick,->] (t7) -- (t10);
    \draw[thick,->] (t10) -- (t8);

    %%%%%%%%%%

    \draw[very thick,->] (Q4) -- (Q5);
    \node at (5,-.5) {(10)8647(10)};
    \node at (5,-1) {$\m^{(4)}$};
    
\end{tikzpicture}
}
% \fi
\caption{Mutation sequence of $\CalQ_4[\{2,3,4\}]$ chain.} 
\label{fig:SU5+3F-open}
\end{figure}

The quiver for $\CalQ_4=\CalQ_4[\{2,3,4\}]$ and the mutations $\m^{(r)}$, $r=1,2,3,4$, are shown in \cref{fig:SU5+3F-open}. 

The first \emph{follow the arrow rule} sequence of mutation $\m^{(1)}$ gives the Baxter $\bQ$-operator and its Fourier transform $Q$-function by:
\begin{align*}
%\begin{split}
    \bQ_4(u) 
    & = \varphi(p_1+u) \varphi(x_2-x_1+u-m_2) \varphi(x_2-x_1+p_2+u) \varphi(p_2+u) \\
    & \quad \times \varphi(x_3-x_2+u-m_3) \varphi(x_3-x_2+p_3+u) \varphi(p_3+u)  \\
    & \quad \times \varphi(x_4-x_3+u-m_4) \varphi(x_4-x_3+p_4+u) \varphi(p_4+u)  \\
    %%%%%
    Q_4(u) & = \int \frac{d\d_1 d\d_2 d\d_3}{(2\pi i)^3} e^{2\pi i (a_1\d_1+a_2\d_2+a_3\d_3)} \bQ_4(u)1 \\
    & = e^{-2\pi i u(a_1+a_2+a_3)} \varphi(u)\varphi(u+a_1) \varphi(u+a_2)\varphi(u+a_3) \\
    & \quad \times e^{-2\pi i (a_1a_2+a_2a_3)} \varphi(a_1+c_b)\varphi(a_2+c_b)\varphi(a_3+c_b) \\
    & \quad \times \frac{\varphi(-a_2-m_2)}{\varphi(a_1-a_2-m_2)} \frac{\varphi(-a_3-m_3)}{\varphi(a_2-a_3-m_3)} \frac{\varphi(-m_4)}{\varphi(a_3-m_4)}
%\end{split}
\end{align*}
The Lax matrices are given by 
\begin{align*}
\begin{split}
    & L_1(u) = \begin{pmatrix}
        1 + e^{2\pi b(p_1+u)} & e^{2\pi b(p_1+x_1+u)}  \\
        e^{-2\pi b x_1} & 0
    \end{pmatrix}~,~ \\
    & L_k(u) = \begin{pmatrix}
        1 + e^{2\pi b(p_k+u)} & e^{2\pi b(p_k+x_k+u)} + e^{2\pi b(x_k+u-m_{k})} \\
        e^{-2\pi b x_k} & e^{2\pi b(u-m_{k})}
    \end{pmatrix} , \quad k=2,3,4,
\end{split}
\end{align*}
and we have the transfer matrix,
\begin{align*}
    \mathring\bt_4(u) = \bQ_4(u-ib/2)\bQ_4(u+ib/2)^{-1} = 1 + \sum_{k=1}^4 H_{4,k} e^{2\pi bku}.
\end{align*}
The Hamiltonians are given by
\begin{subequations}
\begin{align*}
    H_{4,1} = &~ \sum_{j=1}^4 \a_j + \sum_{j=2}^4 (\b_j+\g_j) ,\\
    %%%%%
    H_{4,2} = &~ \sum_{j>k} \a_j\a_k + \left( \sum_{k>j+1} + \sum_{j<k} \right) \a_j(\b_k+\g_k)  + \b_2\b_4 + \sum_{j>k} \g_j\g_k  \\
    & + \b_2\g_4 +\b_3\g_2 + \b_4(\g_2+\g_3) \nonumber ,\\
    %%%%%%
    H_{4,3} = &~ \sum_{j>k>l} \a_j\a_k\a_l + \a_1\a_2(\b_4+\g_4) + \a_3\a_4(\b_2+\g_2) + \a_1\a_4(\b_3+\g_3)  \\
    & + \a_1\g_3\g_4+\a_4\g_2\g_3+\b_4\g_2\g_3 + \g_2\g_3\g_4 \nonumber,\\
    %%%%%%
    H_{4,4} = & ~ \a_1\a_2\a_3\a_4.
\end{align*}
\end{subequations}

\section{Five-dimensional supersymmetric gauge theories}\label{sec:SUSY}

We consider five-dimensional $\CalN=1$ $SU(N)$ supersymmetric gauge theory on non-commutative $\Omega$-deformed $\BR^4_{\bve}\times S^1_\b$~\cite{Nekrasov:2010ka}, where $\b$ is the circumference of compactified $S^1$, with $N_f$ fundamental matters $\bm= (m_1,\dots,m_{N_f})$ and $N_{af}$ anti-fundamental matters $\widetilde\bm=(\widetilde{m}_1,\dots,\widetilde{m}_{N_{af}})$. The Lagrangian is characterized by a complexified gauge coupling 
\begin{align*}
    \kq = e^{2\pi i \tau}, \quad \tau = \frac{4\pi i}{g^2} + \frac{\theta}{2\pi}, \quad \operatorname{Im}\t>0. 
\end{align*}
On the Coulomb branch where the adjoint scalar is given non-vanishing vacuum expectation value, the gauge group is broken to its maximal torus. The vacuum is characterized by the Coulomb moduli parameters $\ba=(a_1,\dots,a_N)$, $ \underline{\ba} = \sum_{i=1}^N a_i = 0$.
Here we denote the $\Omega$-background parameters $\bve=(\ve_1,\ve_2)$ and their exponentiated form $q_i = e^{-\b\ve_i}$.

The perturbative contribution of the partition function is given by%
\footnote{For later convenience, we also include overall constant terms involving only the (anti)-fundamental masses that are usually omitted in the literature.}
\begin{align}
\begin{split}
    & \CalZ^{\rm pert}_{\BR^4 \times S^1} (\ba,\bm,\widetilde{\bm};\bve) \\
    & = \frac{\prod\limits_{1\leq \a\neq \a'\leq N} (e^{\b(a_\a-a_{\a'})};q_1,q_2)_\infty \prod\limits_{1\leq j\neq j'\leq N_f} (e^{\b(m_j-m_{j'})};q_1,q_2)_\infty \prod\limits_{1\leq l\neq l'\leq N_{af}} (e^{\b(\widetilde{m}_l-\widetilde{m}_{l'})};q_1,q_2)_\infty }{\left( \prod\limits_{\a=1}^N \prod\limits_{j=1}^{N_f} (e^{\b(a_\a-m_j)};q_1,q_2)_\infty  \right)
    \left( \prod\limits_{\a=1}^N \prod\limits_{l=1}^{N_{af}} (e^{\b(\widetilde{m}_l-a_\a)};q_1,q_2)_\infty  \right)
    \left( \prod\limits_{j=1}^{N_f} \prod\limits_{l=1}^{N_{af}} (e^{\b(m_j-\widetilde{m}_l)};q_1,q_2)_\infty  \right)}
\end{split}
\end{align}
In the weak coupling limit $g \to 0$, i.e., $\qe \to 0$, which eliminates all the non-perturbative contribution to the partition function. We identify the full partition function as
\begin{align*}
    \CalZ_{\BR^4\times S^1}(\ba,\bm,\widetilde\bm;\bve,\b) = \CalZ^\text{pert}_{\BR^4\times S^1}(\ba,\bm,\widetilde\bm;\bve,\b). 
\end{align*}

\subsection{BPS quiver}

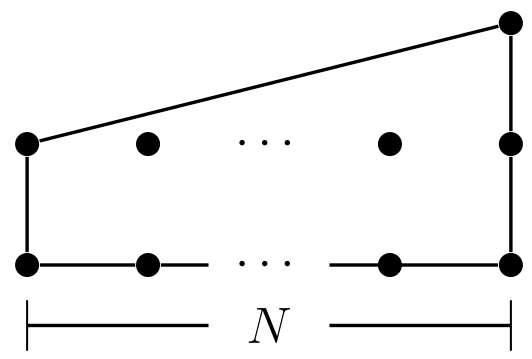
\begin{figure}[h!] 
% \if0
\centering
\scalebox{.8}{
\begin{tikzpicture}[square/.style={regular polygon,regular polygon sides=4},decoration={markings, 
    mark= at position 0.5 with {\arrow[scale=1.3]{stealth}}}
 ]
        
    %%% Toric %%%

    %%% Nodes %%%

    \node at (4,1) [circle,fill,inner sep=2pt] (n11) {};
    \node at (0,0) [circle,fill,inner sep=2pt] (n21) {};
    \node at (1,0) [circle,fill,inner sep=2pt] (n22) {};
    \node at (2,0) (n23) {$\cdots$};
    \node at (3,0) [circle,fill,inner sep=2pt] (n24) {};
    \node at (4,0) [circle,fill,inner sep=2pt] (n25) {};

    \node at (0,-1) [circle,fill,inner sep=2pt] (n31) {};
    \node at (1,-1) [circle,fill,inner sep=2pt] (n32) {};
    \node at (2,-1) (n33) {$\cdots$};
    \node at (3,-1) [circle,fill,inner sep=2pt] (n34) {};
    \node at (4,-1) [circle,fill,inner sep=2pt] (n35) {};

    \draw[thick] (1.5,-1) -- (n32) -- (n31) -- (n21) -- (n11) -- (n25) -- (n35) -- (2.5,-1);

    \node at (0,-1.5) (left) {$|$};
    \node at (4,-1.5) (right) {$|$};
    \node at (2,-1.5) {$N$};
    \draw [thick] (0,-1.5) -- (1.5,-1.5); 
    \draw [thick] (4,-1.5) -- (2.5,-1.5); 
    
\end{tikzpicture}
}
% \fi
\caption{Toric diagram of 5d $\CalN=1$ $U(N)$ SQCD with $N$ fundamental flavors.} 
\label{fig:toric}
\end{figure}

Let us consider a five-dimensional $\mathcal{N}=1$ $U(N)_0$ SQCD with $N$ fundamental matter $\bm=(m_1,\dots,m_N)$ and zero Chern-Simons level on $\BR_\varepsilon^4 \times S_\beta^1$. The theory is characterized by a gauge coupling $\t$, $\operatorname{Im}\t>0$. The corresponding Seiberg-Witten curve is given by 
\begin{align*}
    \S_{\rm SW} = \left\{ (x,y) \in (\BC^\times)^2 \mid x^Ny + \frac{\qe}{y} \prod_{f=1}^N (1- xe^{-m_f}) = \sum_{k=0}^N h_k x^k \right\} .
\end{align*} 
We call the convex Newton polygon of the Seiberg-Witten curve the toric diagram of the 5d $\CalN=1$ supersymmetric gauge theory. See \cref{fig:toric} for illustration.   

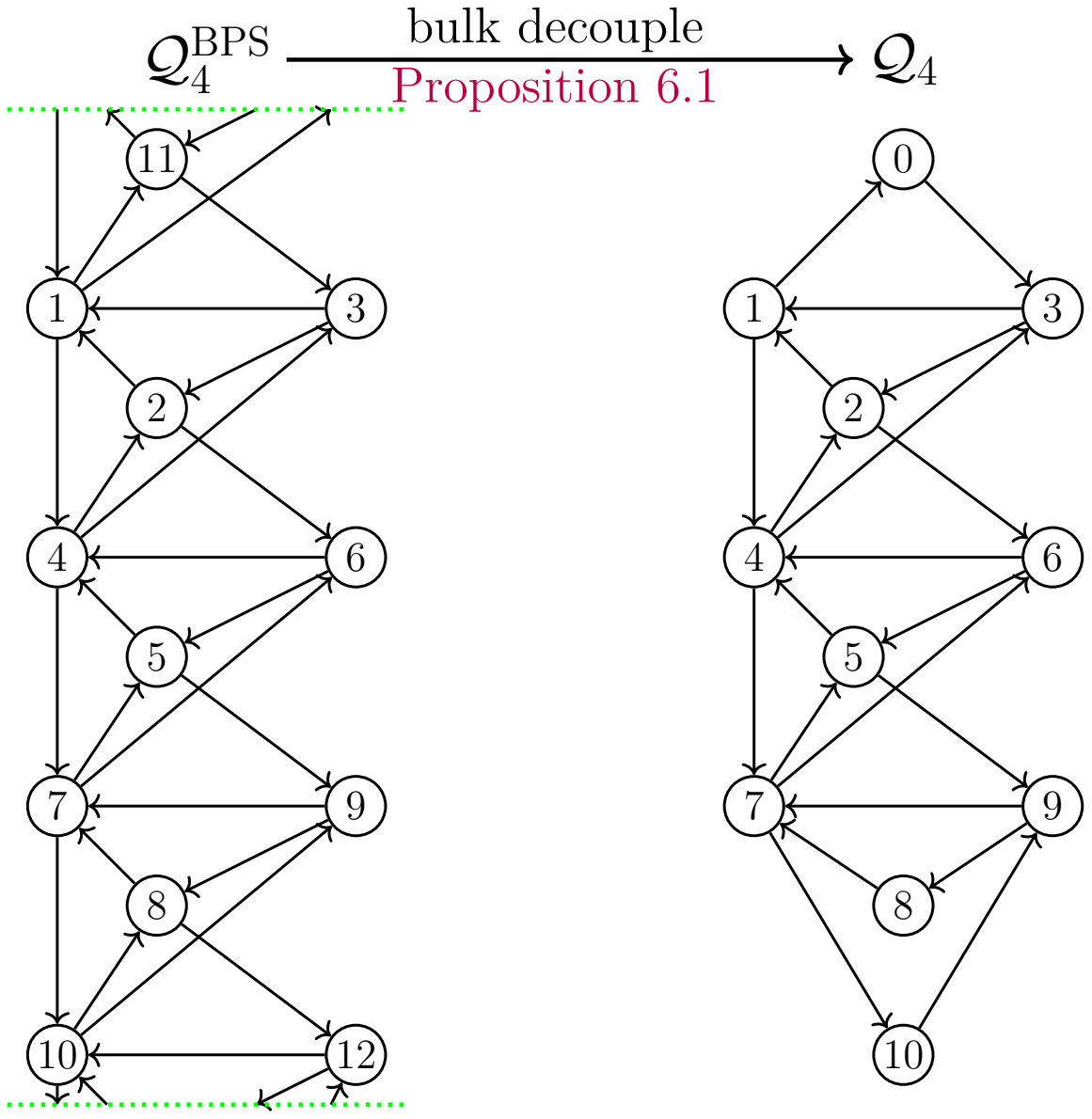
\begin{figure}[h!] 
% \if0
\centering
\scalebox{.7}{
\begin{tikzpicture}[square/.style={regular polygon,regular polygon sides=4},decoration={markings, 
    mark= at position 0.5 with {\arrow[scale=1.3]{stealth}}}
 ]

    %%%%% QUIVER %%%%%
    
    \node at (-3,1.5) [style={circle,draw},thick,inner sep=6pt] (r0) {}; 
    \node at (-4,-0) [style={circle,draw},thick,inner sep=6pt] (r1) {};
    \node at (-1,-0) [style={circle,draw},thick,inner sep=6pt] (r2) {};
    \node at (-3,-1) [style={circle,draw},thick,inner sep=6pt] (r3) {};
    
    \node at (-4,-2.5) [style={circle,draw},thick,inner sep=6pt] (r4) {};
    \node at (-1,-2.5) [style={circle,draw},thick,inner sep=6pt] (r5) {};
    \node at (-3,-3.5) [style={circle,draw},thick,inner sep=6pt] (r6) {};

    \node at (-4,-5) [style={circle,draw},thick,inner sep=6pt] (r7) {};
    \node at (-1,-5) [style={circle,draw},thick,inner sep=6pt] (r8) {};
    \node at (-3,-6) [style={circle,draw},thick,inner sep=6pt] (r9) {};

    \node at (-4,-7.5) [style={circle,draw},thick,inner sep=6pt] (r10) {};
    \node at (-1,-7.5) [style={circle,draw},thick,inner sep=6pt] (r11) {};
    % \node at (-3,-8.5) [style={circle,draw},thick,inner sep=6pt] (r12) {};
    
    % \node at (-4,-10) [style={circle,draw},thick,inner sep=6pt] (r13) {};
    % \node at (-1,-10) [style={circle,draw},thick,inner sep=6pt] (r14) {};
    % \node at (-3,-11) [style={circle,draw},thick,inner sep=6pt] (r15) {};

    \node at (-2.5,2.5) (Q1){\Large $\CalQ_4^{\rm BPS}$};
    \node at (r0) {11};
    \node at (r1) {1};
    \node at (r2) {3};
    \node at (r3) {2};
    \node at (r4) {4};
    \node at (r5) {6};
    \node at (r6) {5};
    \node at (r7) {7};
    \node at (r8) {9};
    \node at (r9) {8};
    \node at (r10) {10};
    \node at (r11) {12};

    %%%%% Arrow %%%%%%

    \draw[thick,->] (-4,2) -- (r1);
    \draw[thick,->] (-2,2) -- (r0);
    \draw[thick,->] (r0) -- (-3.5,2);
    \draw[thick,->] (r1) -- (-1.25,2);

    \draw[thick,->] (r1) -- (r0);
    \draw[thick,->] (r0) -- (r2);

    \draw[thick,->] (r1) -- (r4);
    \draw[thick,->] (r2) -- (r1);
    \draw[thick,->] (r2) -- (r3);
    \draw[thick,->] (r3) -- (r1);
    \draw[thick,->] (r3) -- (r5);
    \draw[thick,->] (r4) -- (r2);
    \draw[thick,->] (r4) -- (r3);
    
    \draw[thick,->] (r4) -- (r7);
    \draw[thick,->] (r5) -- (r4);
    \draw[thick,->] (r5) -- (r6);
    \draw[thick,->] (r6) -- (r4);
    \draw[thick,->] (r6) -- (r8);
    \draw[thick,->] (r7) -- (r5);
    \draw[thick,->] (r7) -- (r6);

    \draw[thick,->] (r7) -- (r10);
    \draw[thick,->] (r8) -- (r7);
    \draw[thick,->] (r8) -- (r9);
    \draw[thick,->] (r9) -- (r7);
    \draw[thick,->] (r9) -- (r11);
    \draw[thick,->] (r10) -- (r8);
    \draw[thick,->] (r10) -- (r9);

    \draw[thick,->] (r10) -- (-4,-8);
    \draw[thick,->] (r11) -- (r10);

    \draw[thick,->] (-1.25,-8) -- (r11);
    \draw[thick,->] (-3.5,-8) -- (r10);
    % \draw[thick,->] (r10) -- (-1.5,-8);
    \draw[thick,->] (r11) -- (-2,-8);

    % \draw[thick,->] (r7) -- (r10);
    % \draw[thick,->] (r10) -- (r8);

    \draw[very thick,green,dotted] (-4.5,2) -- (-.5,2);
    \draw[very thick,green,dotted] (-4.5,-8) -- (-.5,-8);

    %%%%%%%%

    %%%%% QUIVER %%%%%
    
    \node at (4.5,1.5) [style={circle,draw},thick,inner sep=6pt] (q0) {}; 
    \node at (3,-0) [style={circle,draw},thick,inner sep=6pt] (q1) {};
    \node at (6,-0) [style={circle,draw},thick,inner sep=6pt] (q2) {};
    \node at (4,-1) [style={circle,draw},thick,inner sep=6pt] (q3) {};
    
    \node at (3,-2.5) [style={circle,draw},thick,inner sep=6pt] (q4) {};
    \node at (6,-2.5) [style={circle,draw},thick,inner sep=6pt] (q5) {};
    \node at (4,-3.5) [style={circle,draw},thick,inner sep=6pt] (q6) {};

    \node at (3,-5) [style={circle,draw},thick,inner sep=6pt] (q7) {};
    \node at (6,-5) [style={circle,draw},thick,inner sep=6pt] (q8) {};
    \node at (4.5,-6) [style={circle,draw},thick,inner sep=6pt] (q9) {};

    \node at (4.5,-7.5) [style={circle,draw},thick,inner sep=6pt] (q10) {};
    % \node at (4,-7.5) [style={circle,draw},thick,inner sep=6pt] (q11) {};
    % \node at (-3,-8.5) [style={circle,draw},thick,inner sep=6pt] (r12) {};
    
    % \node at (-4,-10) [style={circle,draw},thick,inner sep=6pt] (r13) {};
    % \node at (-1,-10) [style={circle,draw},thick,inner sep=6pt] (r14) {};
    % \node at (-3,-11) [style={circle,draw},thick,inner sep=6pt] (r15) {};

    \node at (4.5,2.5) (Q2){\Large $\mathcal{Q}_4$};
    \node at (q0) {0};
    \node at (q1) {1};
    \node at (q2) {3};
    \node at (q3) {2};
    \node at (q4) {4};
    \node at (q5) {6};
    \node at (q6) {5};
    \node at (q7) {7};
    \node at (q8) {9};
    \node at (q9) {8};
    \node at (q10) {10};
    % \node at (q11) {12};

    %%%%% Arrow %%%%%%

    \draw[thick,->] (q1) -- (q0);
    \draw[thick,->] (q0) -- (q2);

    \draw[thick,->] (q1) -- (q4);
    \draw[thick,->] (q2) -- (q1);
    \draw[thick,->] (q2) -- (q3);
    \draw[thick,->] (q3) -- (q1);
    \draw[thick,->] (q3) -- (q5);
    \draw[thick,->] (q4) -- (q2);
    \draw[thick,->] (q4) -- (q3);
    
    \draw[thick,->] (q4) -- (q7);
    \draw[thick,->] (q5) -- (q4);
    \draw[thick,->] (q5) -- (q6);
    \draw[thick,->] (q6) -- (q4);
    \draw[thick,->] (q6) -- (q8);
    \draw[thick,->] (q7) -- (q5);
    \draw[thick,->] (q7) -- (q6);

    \draw[thick,->] (q7) -- (q10);
    \draw[thick,->] (q8) -- (q7);
    \draw[thick,->] (q8) -- (q9);
    \draw[thick,->] (q9) -- (q7);
    % \draw[thick,->] (q9) -- (q11);
    \draw[thick,->] (q10) -- (q8);
    % \draw[thick,->] (q10) -- (q9);

    %%%%%%

    \draw[very thick,->] (Q1) -- (Q2);
    \node at (1,2.8) {\large bulk decouple };
    \node at (1,2.2) {\large \cref{def:BPS-bulk-decouple}};
    % \node at (1,2.2) {\Large merging $v_{12}$ to $v_{10}$. };
    % \node at (1,1.8) {\Large renaming $v_{11}$ to $v_{0}$. };

\end{tikzpicture}
}
% \fi
\caption{The BPS quiver $\CalQ^{\rm BPS}_4$ of 5d $\CalN=1$ $U(4)$ + $4$ fundamental matter. In the bulk decoupling limit the BPS quuiver $\CalQ^{\rm BPS}_4$ becomes $\CalQ_4$ in \cref{fig:Q3}. } 
\label{fig:BPS}
\end{figure}

Its BPS quiver is constructed using the \emph{fast inverse algorithm} \cite{Closset:2019juk} through its toric diagram. 
We would like to point out that the BPS quiver of 5d $\CalN=1$ SCFT has a superpotential associated to it. We omit the discussion about the superpotential since it does not play any role in the structure of the cluster integrable system. For $N\geq 2$, the BPS quiver $\CalQ_{N}^{\rm BPS}$ for 5d $\CalN=1$ $U(N)$ SQCD, Chern-Simons level zero, with $N$ fundamental flavors is a periodic quiver with $3N$ vertices $V=\{v_1,\dots,v_{3N}\}$, $v_{k+3N}=v_{k}$, and the following set of arrows: 
\[
A = \left\{ \substack{ \displaystyle \{v_{3k} \to v_{3k-1}, v_{3k} \to v_{3k-2}, v_{3k-1} \to v_{3k-2} \}_{k=1}^N , \\ \displaystyle \{ v_{3k-2} \to v_{3k-3}, v_{3k-2} \to v_{3k+1}, v_{3k-1} \to v_{3k+3}, v_{3k+1} \to v_{3k-2} \}_{k=1}^N } \right\} .
\]
See \cref{fig:BPS} for the case $N=4$.

\begin{Proposition}\label{def:BPS-bulk-decouple} 
    Taking the bulk decoupling limit $\kq \to 0$ in the 5d $\CalN=1$ $U(N)$ SQCD with $N$ fundamental flavors modifies the BPS quiver $\CalQ^{\rm BPS}_{N}$ by the following steps:
\begin{enumerate}
    \item merging vertex $v_{3N}$ with vertex $v_{3N-2}$, and 
    \item renaming $v_{3N-1}$ to $v_0$.  
\end{enumerate}
In the bulk decoupling limit the BPS quiver $\CalQ_N^{\rm BPS}$ becomes $\CalQ_N$ defined in \cref{def:quiver}.
\end{Proposition}

We remark that the merged vertex in $\CalQ^{\rm BPS}_N$ becomes the end vertex in $\CalQ_N$. The vertex $v_{3N-1}$, which represents the last fundamental flavor, instead becomes the top vertex in $\CalQ_N$. 

\subsection{Five-dimensional theory on squashed sphere}

We now consider placing the five-dimensional $\CalN=1$ SQCD on a squashed 5-sphere $S^5_{\bom}$ with three parameters $\bom=(\o_1,\o_2,\o_3)$, $\o_i \geq 0$. 

The path integral of 5d $\CalN=1$ theory on a squashed 5-sphere $S^5_{\bom}$ localized to the three fix points of the $U(1)^3$ action on the sphere. The non-perturbative part of the $S^5$ partition function integrand is a triple product of the Nekrasov partition function on the flat space $\BR^4 \times S^1$ with the $\O$-deformation parameters associated to the squashed parameters $\bom$. 
A convenient choice of the local $\O$-parameters are 
\begin{align*}
    \ve_1^{(j)} = \o_k~, \quad \ve_2^{(j)} = \o_l~, \quad \b_j = \frac{2\pi}{\o_j}, \quad (j,k,l) = (1,2,3),~(2,3,1),~(3,1,2).
\end{align*}
The perturbative partition function integrant on a squashed sphere $S^5$ is given by~\cite{Lockhart:2012vp,Imamura:2012efi}% 
\footnote{We also include overall constant terms involving only the (anti)-fundamental masses that are usually omitted in the literature.}
\begin{align*}
\begin{split}
    & \CalZ_{S^5}^{\rm pert}(\ba,\bm,\widetilde\bm;\bom) \\
    & = \frac{\prod\limits_{1\leq \a\neq\b\leq N} \CalS_3(ia_\a - ia_\b|\bom) \prod\limits_{1\leq j\neq j'\leq N_f} \CalS_3(im_j - im_{j'}|\bom) \prod\limits_{1\leq l\neq l'\leq N_{af}} \CalS_3(i\widetilde{m}_l - i\widetilde{m}_{l'}|\bom) }{\left(\prod\limits_{\a=1}^N\prod\limits_{j=1}^{N_f} \CalS_3\left( ia_\a - im_j | \bom\right) \right) \left(\prod\limits_{\a=1}^N\prod\limits_{l=1}^{N_{af}} \CalS_3 \left( i\widetilde{m}_{l} - ia_\a | \bom \right) \right) \left( \prod\limits_{j=1}^{N_f}\prod\limits_{l=1}^{N_{af}} \CalS_3 (im_j-i\widetilde{m}_l+i\o_3|\bom) \right)  } .
\end{split}
\end{align*}
Here $\CalS_3(z|\bom)$ is the triple sine function defined in \eqref{def:multi-sine} in \cref{sec:functions}. 
Then, the partition function is obtained by integration over the Coulomb moduli space: 
\begin{align*}
    \EZ_{S^5}^{\rm pert}(\bm,\widetilde\bm;\bom)  = \int \CalZ_{S^5}^{\rm pert}(\ba,\bm,\widetilde\bm;\bom) d\ba .
\end{align*}
We again consider the weak coupling limit which eliminates all the non-perturbative contribution to the partition function. 

\subsection{Q-function as canonical defect on squashed sphere}

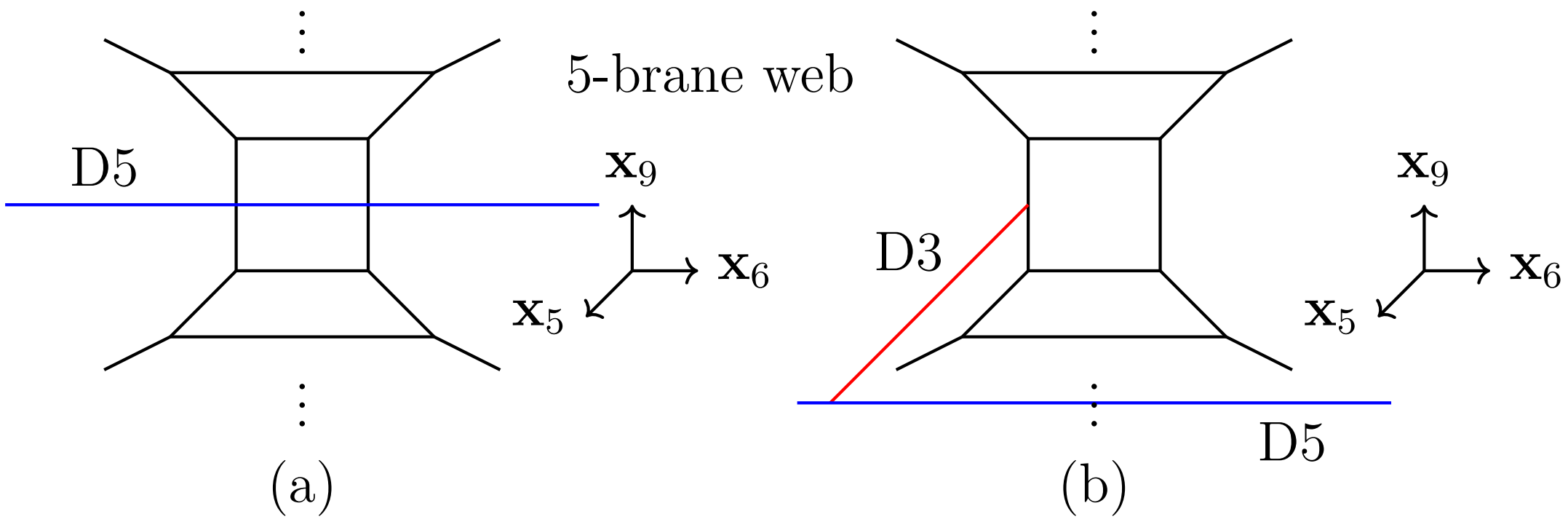
\begin{figure}[h!]
    % \if0
    \centering
    \begin{tikzpicture}[scale=0.6, every node/.style={scale=1.2}]
    \begin{scope}
    \draw[thick] (1,1) -- (-1,1);
    \draw[thick] (1,-1) -- (-1,-1);
    \draw[thick] (2,2) -- (-2,2);
    \draw[thick] (2,-2) -- (-2,-2);
    \draw[thick] (3,2.5) -- (2,2) -- (1,1) -- (1,-1) -- (2,-2) -- (3,-2.5);
    \draw[thick] (-3,2.5) -- (-2,2) -- (-1,1) -- (-1,-1) -- (-2,-2) -- (-3,-2.5);
    \draw[blue,thick] (-4.5,0) -- (4.5,0);

    \node[above] at (0,2) {$\vdots$};
    %\node[left] at (-4,-4) {D3};
    \node[below] at (0,-2) {$\vdots$}; 
    \node[right] at (3.7,2) {5-brane web};
    \node[above] at (-3,0) {D5};

     \draw[->,thick] (5,-1) -- (5,-0);
    \draw[->,thick] (5,-1) -- (6,-1);
    \draw[->,thick] (5,-1) -- (4.3,-1.7);

    \node[above] at (5,0) {$\bx_9$};
    \node[right] at (6,-1) {$\bx_6$};
    \node[left] at (4.3,-1.7) {$\bx_5$};
    
    \node[above] at (0,-5) {(a)};
    
    \end{scope}
    \begin{scope}[xshift = 12cm]
    \draw[thick] (1,1) -- (-1,1);
    \draw[thick] (1,-1) -- (-1,-1);
    \draw[thick] (2,2) -- (-2,2);
    \draw[thick] (2,-2) -- (-2,-2);
    \draw[thick] (3,2.5) -- (2,2) -- (1,1) -- (1,-1) -- (2,-2) -- (3,-2.5);
    \draw[thick] (-3,2.5) -- (-2,2) -- (-1,1) -- (-1,-1) -- (-2,-2) -- (-3,-2.5);
    \draw[red,thick] (-1,0) -- (-4,-3);
    \draw[blue,thick] (-4.5,-3) -- (4.5,-3);

    \node[above] at (0,2) {$\vdots$};
    \node[left] at (-2,-0.7) {D3};
    \node[below] at (0,-2) {$\vdots$}; 
    %\node[right] at (3.5,2) {5-brane web};
    \node[below] at (3,-3) {D5};

    \draw[->,thick] (5,-1) -- (5,-0);
    \draw[->,thick] (5,-1) -- (6,-1);
    \draw[->,thick] (5,-1) -- (4.3,-1.7);

    \node[above] at (5,0) {$\bx_9$};
    \node[right] at (6,-1) {$\bx_6$};
    \node[left] at (4.3,-1.7) {$\bx_5$};

    \node[above] at (0,-5) {(b)};
    \end{scope}
    \end{tikzpicture}
    % \fi
    \caption{D3-brane creation and higgsing construction of $Q$-function. (a) One of the D5-branes is aligned across the two NS5-branes, enabling the D5-brane to move along the $x^5$-direction. (b) A D3-brane ending on one of the NS5-branes is created as a result of the transition.}
    \label{fig:Q-operator}
\end{figure}

Let us consider the 5d $\mathcal{N}=1$ $U(N+1)$ SQCD with $N+1$ fundamental matters and one anti-fundamental matter of the mass $\widetilde{m}$. In the 5-brane web construction, the $U(N+1)$ SQCD has $N+1$ semi-infinite D5 branes on the right and 1 semi-infinite D5 brane on the left as shown in \cref{fig:Q-operator}.

Let us isolate the contribution coming from the cross term between $(N+1)$-th Coulomb moduli $a_{N+1}$, fundamental matter $m_{N+1}$ and the anti-fundamental matter $\widetilde m$, i.e. the parameter in addition to the $U(N)$ SQCD with $N$-fundamental matter:
\[
    \CalZ_{\rm extra} (a_{N+1},m_{N+1},\widetilde m|\bom) = \frac{1}{\CalS_3(ia_{N+1}-im_{N+1}|\bom)\CalS_3(i\widetilde m - ia_{N+1}|\bom) \CalS_3(im_{N+1}-i\widetilde{m}+i\o_3|\bom) }
\]

We impose a Higgsing condition by tuning $a_{N+1}$, $m_{N+1}$ and $\widetilde{m}$:
\[
    a_{N+1}= m_{N+1} = \widetilde{m}-i\o_3 = u-i\frac{\o_1+\o_2}{2}-i\o_3.
\]
In the 5-brane picture, the Higgsing aligns the last finite size D5 brane with the semi-infinite D5 branes on both sides. We can then pull the aligned D5 branes away from the 5-brane web, creating a single D3 brane stretching between the pulled away D5 and the remaining 5-brane web, which results a $U(N)$ gauge theory with $N$ fundamental flavors. See \cref{fig:Q-operator} for illustration.
The D3 brane acts as a co-dimensional two defect in the 5d $\CalN=1$ SQCD.

We consider the following operation for the integrand of the perturbative partition function of the $U(N+1)$ gauge theory with Higgsing \cite{Gaiotto:2014ina}:
\begin{align}
\begin{split}
    & \lim_{\substack{a_{N+1} \to u-i\frac{\o_1+\o_2}{2}-i\o_3 \\
    m_{N+1} \to u-i\frac{\o_1+\o_2}{2}-i\o_3 \\
    \widetilde m \to u-i\frac{\o_1+\o_2}{2}
    }} \CalZ_{\rm extra} ^{-1}(a_{N+1},m_{N+1},\widetilde m;\bom) \CalZ^{U(N+1)}_{S^5}(\ba,a_{N+1},\bm,m_{N+1},\widetilde{m};\bom) \\
    & = \CalZ^{U(N)}_{S^5}(\ba,\bm;\bom) \prod_{\a=1}^N \CalS_2 \left( iu - ia_\a + \frac{\o_1+\o_2}{2} \mid \o_1,\o_2\right)
\end{split}
\end{align}
We define the gauge theory $Q$-function by 
\begin{align*}
    Q_\ba^{\rm gauge}(u;\o_1,\o_2):= \prod_{\a=1}^N \CalS_2 \left( iu - ia_\a + \frac{\o_1+\o_2}{2} \mid \o_1,\o_2\right).
\end{align*}
The D3 brane created from the Higgsing that warps a squashed $S^3_b$.
We choose $\o_1=b$ and $\o_2 = b^{-1}$.  This links the double sine function with the quantum dilogarithm function \eqref{eq:double-sine-qdf}.  The $Q$-function representing the D3 brane becomes:
\begin{align*}
    Q_\ba^{\rm gauge}(u;b,b^{-1}) = \zeta^{N} \prod_{\a=1}^N \varphi(u-a_\a). 
\end{align*}
This $Q$-function is, up to a shift in argument and overall constant, Fourier transform (\cref{def:Q-function}) of the Baxter $\bQ$-operator (\cref{prop:Q-function-quiver}).

\appendix

\section{Special functions}\label{sec:functions}

\subsection{Faddeev's quantum dilogarithm function}
We follow the notation of \cite{kashaev2001quantum}. Set $b \in \mathbb{R}_{>0}$. 
We define
\begin{align*}
    c_b = \frac{i(b+b^{-1})}{2}~,~ \D_b = \frac{i(b-b^{-1})}{2} ~,~ \zeta = e^{\frac{\pi i}{12}(1-4c_b^2)}~,~ \zeta_{\text{inv}} = \zeta^{-2} e^{-\pi i c_b^2}.
\end{align*}

The Faddeev's quantum dilogarithm function is defined on the strip $|\operatorname{Im}(z)|<|\operatorname{Im}(c_b)|$ by the following formula:
\begin{align*}%\label{def:qdf}
\begin{split}
    \varphi(z) & = \exp \left( \frac{1}{4} \int_{\BR+i0} \frac{e^{-2izt}}{\sinh(bt)\sinh(b^{-1}t)} \frac{dt}{t} \right) .
\end{split}
\end{align*}
This function can be analytically continued to the full complex plane with an essential singularity at infinity together with the following properties:
\begin{subequations}
\begin{align}
    \text{Zeros and poles: } & \varphi(z)^{\pm1} =0 \quad \text{at} \quad z = \mp (c_b + ibm + ib^{-1}n)~,~ n,m \in \mathbb{Z}_{\ge 0} \\
    \text{Reflection formula: } & \varphi(z) \varphi(-z) = \zeta_{\text{inv}} e^{\pi i z^2} \label{qdf-inversion} \\
    \text{Functional equations: } & \varphi(z-ib^\pm/2) = \left( 1 + e^{2\pi b^\pm z} \right) \varphi (z+ib^\pm/2) \label{eq:qdf-gamma-like} \\
    \text{Fourier transform: } & \zeta \varphi(w) = \int_{\BR+i0} \frac{e^{2\pi iz(w-c_b)}}{\varphi(z-c_b)} dz~,~ \frac{1}{\zeta\varphi(w)} = \int_{\BR-i0} \frac{\varphi(z+c_b)}{e^{2\pi iz(w+c_b)}} dz \label{eq:qdf-fourier} \\
    \text{Asymptotics: } & \varphi(z) \sim \begin{cases}
        \zeta_{\rm inv} e^{\pi i z^2}, & {\rm Re}(z) \to \infty \\
        1, & {\rm Re}(z) \to -\infty
    \end{cases}
\end{align}
\end{subequations}

For self-adjoint operators $p$ and $x$ satisfying $[p,x] = \frac{1}{2\pi i}$, we have
\begin{subequations}
\begin{gather}
    \varphi(x+ib/2) e^{2\pi b p} = e^{2\pi b p} \varphi(x-ib/2), \label{eq:qdf-passing} \\
    \varphi(x)^{-1} e^{2\pi b p} \varphi(x) = \varphi(p) e^{2\pi b x} \varphi(p)^{-1}=e^{2\pi b x} + e^{2\pi b(p+x)} . \label{eq:qdf-conjugation} \\
    \varphi(p) \varphi(x) = \varphi(x) \varphi(p+x) \varphi(p). \label{eq:qdf-pentagon}
\end{gather}
\end{subequations}
The third relation is known as the pentagon identity, which is equivalent to the following integral formulas~\cite{faddeev2001strongly}:
\begin{align}
\begin{split}
    & \frac{\varphi(a)\varphi(w)}{\varphi(a+w-c_b)} = \zeta^{-1} \int_{\BR+i0} \frac{dx}{2\pi i} \frac{\varphi(x+a)}{\varphi(x-c_b)} e^{2\pi ix(w-c_b)} ,\\
    & \frac{\varphi(a+w+c_b)}{\varphi(a)\varphi(w)} = \zeta \int_{\BR-i0} \frac{dx}{2\pi i} \frac{\varphi(x+c_b)}{\varphi(x+a)} e^{-2\pi ix(w+c_b)}. \label{eq:qdf-Ramanujan}
\end{split}
\end{align}

\subsection{Multiple sine functions}\label{sec:multiple_functions}

We follow the notation in \cite{narukawa2004modular,komori2010barnes}. 
Let 
\[
    \bom = (\o_1,\dots,\o_r)~,~ \o_k >0.
\]
We define the Barns multiple zeta function
\begin{align*}
    \zeta_r(s,z|\bom) = \sum_{n_1,\dots,n_r\geq 0} 
    (z+n_1\o_1+\cdots+n_r\o_r)^{-s} .
\end{align*}
The multiple gamma function is defined from the multiple zeta function through analytic continuation
\begin{align*}
    \G_r(z|\bom) = \exp \left( \left. \frac{\p}{\p s} \zeta_r(s,z|\bom) \right|_{s=0} \right).
\end{align*}
Then, we define the multiple sine function
\begin{align}\label{def:multi-sine}
    \CalS_r(z|\bom) = \G_r(z|\bom)^{-1} \G(\o-z|\bom)^{(-1)^r}
\end{align}
where $\o=\o_1+\cdots+\o_r$. 
In particular, the double sine function is given by
\begin{align*}
    \CalS_2(z|\o_1,\o_2) = \prod_{n_1,n_2\geq 0}^\infty \frac{z+n_1\o_1+n_2\o_2}{\o-z+n_1\o_1+n_2\o_2}
\end{align*}
We may also write
\begin{align*}
    \CalS_2(z|\omega_1,\omega_2) = \frac{(e^{2\pi i z/\omega_1};e^{2\pi i \omega_2/\omega_1})_\infty}{(e^{2\pi i (\omega - z)/\omega_1};e^{2\pi i \omega_2/\omega_1})_\infty} = \frac{(e^{2\pi i z/\omega_2};e^{2\pi i \omega_1/\omega_2})_\infty}{(e^{2\pi i (\omega - z)/\omega_2};e^{2\pi i \omega_1/\omega_2})_\infty} \, ,
\end{align*}
and
\begin{align*}
    \CalS_2(z|\omega_1,\omega_2) = \exp \left( \int_{\mathbb{R}+i0} \frac{e^{izw}}{(e^{\omega_1 w}-1)(e^{\omega_2 w}-1)} \frac{dw}{w} \right) \, .
\end{align*}
Moreover, the double sine function is related to the quantum dilogarithm function as follows, 
\begin{align}\label{eq:double-sine-qdf}
    \CalS_2(-iz|b,b^{-1}) = \zeta \varphi(c_b-z).
\end{align}
The triple sine function is given by
\begin{align*}%\label{eq:triple-sine}
    \CalS_3(z|\o_1,\o_2,\o_3) = z\prod_{\substack{n_1,n_2,n_3 \geq 0\\ n_1+n_2+n_3>0}}^\infty \frac{(z+n_1\o_1+n_2\o_2+n_3\o_3)(\o-z+n_1\o_1+n_2\o_2+n_3\o_3)}{(n_1\o_1+n_2\o_2+n_3\o_3)(\o+n_1\o_1+n_2\o_2+n_3\o_3)}. 
\end{align*}

\section{Flavored Toda chain}\label{sec:flavored_Toda_rational}

\subsection{Rational case}

A rational flavored Toda chain of length $N$ with twisted periodic boundary condition is characterized by a set of flavored parameters $\{m_j\}_{j\in F}$, $F\subset \{1,2,\dots,N\}$, and a twisted parameter $\qe_{\rm rat}$. The system is described by the Hamiltonian:
\begin{align}\label{def:fToda-H}
    H _{\rm fT}^{\rm rat}= \sum_{j=1}^N \frac{p_j^2}{2} + \sum_{j=2}^N e^{x_j-x_{j-1}} \left( 1 + {\bf 1}_F(j) \frac{p_j}{m_j} \right) + \qe_{\rm rat} e^{x_1-x_N} \left( 1 + {\bf 1}_F(1) \frac{p_1}{m_1} \right).
\end{align}
where the canonical variables $\bx=(x_1,\dots,x_N)$ and $\bp=(p_1,\dots,p_N)$ obey the commutation relation
\[
    [x_j,p_k] = i\hbar \d_{jk}. 
\]
A flavor at the $j$-th site can be removed by sending $m_j \to \infty$. The system recovers the ordinary  rational Toda chain if all the flavors are removed. 

A flavored Toda chain is a quantum integrable system. For later convenience, we shift 
\begin{gather*}
    \qe_{\rm rat} \mapsto \qe_{\rm rat} \prod_{j\in F} m_j, \qquad 
    x_j-x_{j-1} \mapsto x_j-x_{j-1} - \log m_j,~ j\in F .
\end{gather*}
The Hamiltonian becomes
\begin{align}\label{eq:fToda-H-m}
    H _{\rm fT}^{\rm rat}= \sum_{j=1}^N \frac{p_j^2}{2} + \sum_{j=2}^N e^{x_j-x_{j-1}} (m_j+p_j)^{{\bf 1}_F(j)} + \qe_{\rm rat} e^{x_1-x_k} (m_1+p_1)^{{\bf 1}_F(1)}.  
\end{align}
The corresponding Lax operators of rank two are 
\begin{align}\label{eq:Lax-rational}
\begin{split}
    & L_j^{\rm rat}(u) = \begin{pmatrix}
        u-p_j & -e^{x_j}(p_j+m_j)^{{\bf 1}_F(j)} \\
        e^{-x_j}& {\bf 1}_F(j)
    \end{pmatrix},~ j=1,\dots,N.  
\end{split}
\end{align}
The flavor parameters can be considered as inhomogeneity at the lattice sites by considering a shift of momentum $p_j \mapsto p_j-{\bf 1}_F(j)m_j$:
\[
    L_j^{\rm rat}(u) = \begin{pmatrix}
        u+{\bf 1}_F(j)m_j-p_j & -e^{x_j}p_j^{{\bf 1}_F(j)} \\
        e^{-x_j}& {\bf 1}_F(j)
    \end{pmatrix},
\]
The monodromy matrix is obtained by concatenating the Lax operators 
\begin{align}
\begin{split}
    & \bT_N^{\rm rat}(u) = \begin{pmatrix}
        1 & 0 \\ 0 & \qe_{\rm rat} 
    \end{pmatrix} L_k^{\rm rat}(u) \cdots L_2^{\rm rat}(u) L_1^{\rm rat}(u), \\
    & \bt_N^{\rm rat}(u) = \Tr \bT_N^{\rm rat}(u) = \sum_{j=0}^N (-1)^j H_{N,j}^{\rm rat} u^{N-j}. 
\end{split}
\end{align}
The defining Hamiltonian of the flavored Toda lattice \eqref{eq:fToda-H-m} is
\[
    H_{\rm fT}^{\rm rat} = \frac{(H_{N,1}^{\rm rat})^2}{2} - H_{N,2}^{\rm rat}.
\]
The open flavored Toda chain is obtained by sending $\qe_{\rm rat} \to 0$:
\begin{align}\label{def:fToda-H-open}
    H_{\rm fT}^{\rm rat}|_{\rm open} = \sum_{j=1}^N \frac{p_j^2}{2} + \sum_{j=2}^N  e^{x_j-x_{j-1}} (m_j+p_j)^{{\bf 1}_F(j)} . 
\end{align}

\subsection{$q$-deformed case}

A rational flavored Toda chain of length $N$ with twisted periodic boundary condition is characterized by a set of flavored parameters $\{m_j\}_{j\in F}$. $F\subset \{1,2,\dots,N\}$ and a twisted parameter $\qe_q$. The system is described by the Hamiltonian:
\begin{align}\label{def:fToda-H-q}
    H _{\rm fT}^{q}= \sum_{j=1}^N e^{p_j} + \sum_{j=2}^N e^{x_j-x_{j-1}} \left( e^{p_j} - {\bf 1}_F(j) e^{-m_j} \right) + \qe_q e^{x_1-x_N} \left( e^{p_1} - {\bf 1}_F(1) e^{-m_1} \right).
\end{align}
The flavors can be removed by sending $m_j \to \infty$. The system recovers the ordinary  $q$-deformed Toda chain when all flavors are removed. 

A $q$-deformed Toda chain is quantum integrable. 
The corresponding Lax operators of rank two are 
\begin{align}\label{eq:Lax-q-deformed}
     L_{j}^q(u)= \begin{pmatrix}
        1 - e^{p_j-u} & -e^{x_j-u} (e^{p_j}-{\bf 1}_F(j) e^{-m_j})  \\
        e^{-x_j} & \mathbf{1}_F(j) e^{-u-m_{j}}
    \end{pmatrix}.
\end{align}
The monodromy matrix is obtained by concatenating the Lax operators 
\begin{align}
\begin{split}
    & \bT_N^{q}(u) = \begin{pmatrix}
        1 & 0 \\ 0 & \qe_q 
    \end{pmatrix} L_N^{q}(u) \cdots L_2^{q}(u) L_1^{q}(u),~ \\
    & \bt_N^{q}(u) = \Tr \bT_N^{q}(u) = \sum_{j=0}^N (-1)^j H_{N,j}^{q} e^{-ju}. 
\end{split}
\end{align}
The defining Hamiltonian of the flavored Toda lattice \eqref{eq:fToda-H-m} is
\[
    H_{\rm fT}^{q} = H_{N,1}^{q}.
\]
The open flavored Toda chain is obtained by sending $\qe_q \to 0$:
\begin{align}\label{def:fToda-H-q-open}
    H _{\rm fT}^{q}= \sum_{j=1}^N e^{p_j} + \sum_{j=2}^N e^{x_j-x_{j-1}} \left( e^{p_j} - {\bf 1}_F(j) e^{-m_j} \right) 
\end{align}

\subsection{From $q$-deformed to rational}\label{sec:q-deform_to_rational}

The rational flavored Toda chain \ref{def:fToda-H} can be obtained from taking non-relativistic limit from the $q$-deformed flavored Toda chain \eqref{def:fToda-H-q} obtained from $\CalQ_N[F]$ through the following steps:
\begin{enumerate}
    \item Restore the dependence on the speed of light $c$ by $p_j \mapsto \frac{p_j}{c}$, $m_j \mapsto \frac{m_j}{c}$, $u \mapsto \frac{u}{c}$ and shift $x_j \mapsto x_j +(1-d_j)\log c$, 
    $d_j = \sum_{l=1}^{j} (2-{\bf 1}_F(l))$. 
    \item Take the following transformation for the Lax operators:
    \[
        L_j(u) \mapsto \begin{pmatrix}
            1 & 0 \\ 0 & c^{-d_{j}} 
        \end{pmatrix} L_j(u) \begin{pmatrix}
            1 & 0 \\ 0 & c^{d_{j-1}} 
        \end{pmatrix},~ j=1,\dots,N.
    \]
    \item Identify the relation between the rational and $q$-deformed twisted parameters $\qe_q = \qe_{\rm rat} c^{-d_N}$. 
    \item Send $c \gg 1$ and expand $H_{N,1}^q$ and $L_j^q(u)$ in orders of $c^{-1}$. The $c^{-2}$ order in $H_{N,1}^q$ and $c^{-1}$ order in $L_j^q(u)$ gives the rational open flavored Toda chain Hamiltonian \eqref{def:fToda-H-open} and its Lax operators \eqref{eq:Lax-rational}.
\end{enumerate}

\section{Proofs}

\subsection{Proof of \cref{lem:T-as-operator}}\label{proof:lem:T-as-operator}

It is sufficient to proof the case where $k\in F$. 
    Applying \eqref{eq:qdf-fourier} twice, we obtain
    \begin{align}\nonumber
    \begin{split}
        T_k(u) f(\bx)
        & = \varphi(x_{k}-x_{k-1}+u-m_k) \zeta^{-2}
        \int_{\mathbb{R}+i0} \frac{e^{2\pi i t_1(p_k+x_k-x_{k-1}+u-c_b)}}{\varphi(t_1-c_b)} \frac{e^{2\pi it_2(p_k+u-c_b)}}{\varphi(t_2-c_b)} f(\bx) dt_1dt_2 \\
        & = \varphi(x_{k}-x_{k-1}+u-m_k) \zeta^{-2}
        \int_{\mathbb{R}+i0} \frac{e^{2\pi i t_1(x_k-x_{k-1}+u-c_b)}e^{\pi i t_1^2}}{\varphi(t_1-c_b)} \frac{e^{2\pi it_2(u-c_b)}}{\varphi(t_2-c_b)} e^{2\pi i (t_1+t_2)p_k} f(\bx) dt_1dt_2 \\
        & = \varphi(x_{k}-x_{k-1}+u-m_k) \zeta^{-2}
        \int_{\mathbb{R}+i0}  \frac{e^{2\pi i t_1(x_k-x_{k-1})}e^{\pi i t_1^2}e^{2\pi i (t_1+t_2)(u-c_b)}}{\varphi(t_1-c_b)\varphi(t_2-c_b)} f(\bx)|_{x_k\to x_k+t_1+t_2} dt_1dt_2 .
    \end{split}
    \end{align}
    We write $y_k=x_k+t_1+t_2$. Reparametrizing $t_1 \mapsto -t_1$  and using the reflection formula \eqref{qdf-inversion}, we obtain
    \begin{align}\nonumber
    \begin{split}
        T_k(u) f(\bx)
        & = \varphi(x_{k}-x_{k-1}+u-m_k) \zeta^{-2}
        \int_{\mathbb{R}-i0}  \frac{e^{-2\pi i t_1(x_k-x_{k-1})}e^{2\pi i (y_k-x_k)(u-c_b)}\varphi(t_1+c_b)}{\varphi(y_k-x_k+t_1-c_b)} f(\bx)|_{x_k\to y_k} dt_1dy_k
    \end{split}
    \end{align}
    Finally, using \eqref{eq:qdf-Ramanujan} to take the integration on $t_1$ gives
    \begin{align}\nonumber
        T_k(u) f(\bx) = \varphi(x_{k}-x_{k-1}+u-m_k) \zeta^{-1} \int_{\mathbb{R}+i0}  \frac{e^{2\pi i(y_k-x_k)(u-c_b)}\varphi(y_k-x_{k-1})}{\varphi(y_k-x_k-c_b)\varphi(x_k-x_{k-1})} f(\bx)|_{x_k\to y_k} dy_k. 
    \end{align}
    For $k\notin F$, we simply take the limit $m_k \to \infty$, so that $\varphi(x_{k}-x_{k-1}+u-m_k) \to 1$.
    \hfill $\square$

\subsection{Proof of \cref{prop:Q-function-quiver}}\label{proof:prop:Q-function-quiver}

We first prove the case for $\CalQ_N=\CalQ_N[\{2,\dots,N\}]$. 
    Start with \cref{cor:Baxter-integral} and take Fourier transform with respect to $\d_k=x_{k+1}-x_k$:
    \begin{multline*}
%    \begin{split}
        Q_\ba(u)
        = \int_{(\BR+i0)^{N-1}} \left( \zeta^{1-2N} \int_{(\BR+i0)^N}  e^{2\pi i (\underline{\bx}-\underline{\by})(c_b-u)} \prod_{k=1}^N \frac{1}{\varphi(y_k-x_k-c_b)} \right. \\
        \left. \quad \times \prod_{k=1}^{N-1} \frac{\varphi(x_{k+1}-y_k+u-m_{k+1})\varphi(y_{k+1}-y_k)}{\varphi(x_{k+1}-y_k)}1 d\by \right) \prod_{k=1}^{N-1} e^{2\pi i\d_k a_k} \frac{d\d_k}{2\pi i} 
%    \end{split}
    \end{multline*}
    Writing $y_k = x_k+t_k$ and applying the shift $\d_k \mapsto \d_k+t_k$, we have
    \begin{align*}
    \begin{split}
        & Q_\ba(u) \\
        & = \zeta^{1-2N} \int_{(\BR+i0)^{N-1}} \left( \int_{(\BR+i0)^{N}} \prod_{k=1}^N \frac{e^{2\pi i t_k(u-c_b)}}{\varphi(t_k-c_b)} % \\ & \quad \times 
        \prod_{k=1}^{N-1} \frac{\varphi(\d_k-t_k+u-m_k)\varphi(\d_k+t_{k+1}-t_k)}{\varphi(\d_k-t_k)} d\bt \right) \prod_{k=1}^{N-1} e^{2\pi i\d_k a_k} \frac{d\d_k}{2\pi i} \\
        & = \zeta^{1-2N} \int_{(\BR+i0)^{2N-1}} \prod_{k=1}^{N-1} \frac{e^{2\pi i t_k(u+a_k-c_b)}}{\varphi(t_k-c_b)} \frac{\varphi(\d_k+u-m_{k+1})\varphi(\d_k+t_{k+1})}{\varphi(t_k-c_b)\varphi(\d_k)} \frac{e^{2\pi i t_N(u-c_b)}}{\varphi(t_{N}-c_b)} e^{2\pi i\d_k a_k} \frac{d\d_k}{2\pi i} d\bt .
    \end{split}
    \end{align*}
    We integrate out $t_N$ using \eqref{eq:qdf-Ramanujan}, then move on to the integration for $\d_{N-1}$ and $t_{N-1}$: 
    \begin{align*}
    \begin{split}
        Q_\ba(u)
        & = \zeta^{2-2N} \int_{\BR+i0} \prod_{k=1}^{N-2} \frac{d\d_kdt_k}{2\pi i}  e^{2\pi i\d_k a_k} \frac{e^{2\pi i t_k(u+a_k-c_b)}}{\varphi(t_k-c_b)} \frac{\varphi(\d_k+u-m_{k+1})\varphi(\d_{k-1}+t_{k})^{\d_{k>1}}}{\varphi(t_k-c_b)\varphi(\d_k)} \\
        & \quad \times \int_{\BR+i0} \frac{d\d_{N-1}dt_{N-1}}{2\pi i} e^{2\pi i \d_{N-1}a_{N-1}} e^{2\pi i t_{N-1}(u+a_{N-1}-c_b)} \frac{\varphi(\d_{N-1}+u-m_{N})\varphi(\d_{N-2}+t_{N-1})}{\varphi(t_{N-1}-c_b)\varphi(\d_{N-1})} \\
        & \quad \times 2\pi i \frac{\varphi(\d_{N-1})\varphi(u)}{\varphi(\d_{N-1}+u-c_b)} \\
        %%%%%%%%%%
        & = \varphi(u) \zeta^{4-2N}\int_{\BR+i0} \prod_{k=1}^{N-1} \frac{d\d_kdt_k}{2\pi i}  e^{2\pi i\d_k a_k} \frac{e^{2\pi i t_k(u+a_k-c_b)}}{\varphi(t_k-c_b)} \frac{\varphi(\d_k+u-m_{k+1})\varphi(\d_{k-1}+t_{k})^{\d_{k>1}}}{\varphi(t_k-c_b)\varphi(\d_k)} \\
        & \quad \times \zeta^{-1} \int_{\BR+i0} dt_{N-1} \frac{\varphi(t_{N-1}+\d_{N-2})}{\varphi(t_{N-1}-c_b)} e^{2\pi i t_{N-1}(u+a_{N-1}-c_b)} \\
        & \quad \times \zeta^{-1} \int_{\BR+i0} d\d_{N-1} \frac{\varphi(\d_{N-1}+u-m_N)}{\varphi(\d_{N-1}+u-c_b)} e^{2\pi i \d_{N-1}a_{N-1}} 
    \end{split}
    \end{align*}
    Shifting $\d_{N-1} \mapsto \d_{N-1}-u$ and using \eqref{eq:qdf-Ramanujan} for both integrations give
    \begin{align*}
    \begin{split}
        Q_\ba(u)
        & = \varphi(u) \zeta^{4-2N} \int_{\BR+i0} \prod_{k=1}^{N-2} \frac{d\d_kdt_k}{2\pi i}  e^{2\pi i\d_k a_k} \frac{e^{2\pi i t_k(u+a_k-c_b)}}{\varphi(t_k-c_b)} \frac{\varphi(\d_k+u-m_{k+1})\varphi(\d_{k-1}+t_{k})^{\d_{k>1}}}{\varphi(t_k-c_b)\varphi(\d_k)} \\
        & \quad \times 2\pi i\frac{\varphi(\d_{N-2})\varphi(u+a_{N-1})}{\varphi(u+a_{N-1}+\d_{N-2}-c_b)} e^{-2\pi i ua_{N-1}} \frac{\varphi(-m_{N})\varphi(a_{N-1}+c_b)}{\varphi(a_{N-1}-m_N-c_b)} 
    \end{split}
    \end{align*}
    Repeating the the integration gives the result. 
    For $\CalQ_N[F]$ with $F\subsetneq\{2,\dots,N\}$, we simply take the limit $m_k \to \infty$ if $k\notin F$.
    \hfill $\square$

\subsection{Proof of \cref{lem:TT}} \label{proof:lem:TT}

It is enought to prove the case for $\CalQ_N$. 
We shift ${p}_k \mapsto p_k+m_k$ and $u\mapsto u-m_k$, $v \mapsto v-m_k$ to shorten the computation as follows:
\begin{align}\nonumber
\begin{split}
    & T_k(u)T_k(v) \\
    & = \varphi(x_k-x_{k-1}+u) \varphi(p_k+x_k-x_{k-1}+u)\varphi(p_k+u) \varphi(x_k-x_{k-1}+v) \varphi(p_k+x_k-x_{k-1}+v)\varphi(p_k+v) \\
    & = \varphi(x_k-x_{k-1}+u) \varphi(p_k+x_k-x_{k-1}+u) \varphi(x_k+v) \varphi(p_k+x_k-x_{k-1}+u+v) \varphi(p_k+u) \\
    & \qquad \varphi(p_k+x_k-x_{k-1}+v)\varphi(p_k-x_{k-1}+v) \\
    & = \varphi(x_k-x_{k-1}+u)  \varphi(x_k-x_{k-1}+v) \varphi(p_k+2x_k-2x_{k-1}+u+v) \varphi(p_k+x_k-x_{k-1}+u)  \\
    & \qquad \varphi(p_k+x_k-x_{k-1}+u+v)\varphi(p_k+x_k-x_{k-1}+v) \varphi(2p_k+x_k-x_{k-1}+u+v) \varphi(p_k+u) \varphi(p_k+v) \\
    %%%%%%%%%%
    & = \varphi(x_k-x_{k-1}+v)  \varphi(x_k-x_{k-1}+u) \varphi(p_k+2x_k-2x_{k-1}+u+v) \varphi(p_k+x_k-x_{k-1}+v) \\
    & \qquad \varphi(p_k+x_k-x_{k-1}+u+v)  \varphi(p_k+x_k-x_{k-1}+u) \varphi(2p_k+x_k-x_{k-1}+u+v) \varphi(p_k+v) \varphi(p_k+u) \\
    %%%%%%%%
    & = \varphi(x_k-x_{k-1}+v) \varphi(p_k+x_k-x_{k-1}+v) \varphi(x_k-x_{k-1}+u) \varphi(p_k+x_k-x_{k-1}+u+v) \varphi(p_k+v) \\
    & \qquad \varphi(p_k+x_k-x_{k-1}+u) \varphi(p_k+u) \\
    & = \varphi(x_k-x_{k-1}+v) \varphi(p_k+x_k-x_{k-1}+v) \varphi(p_k+v) \varphi(x_k-x_{k-1}+u) \varphi(p_k+x_k-x_{k-1}+u) \varphi(p_k+u) \\
    & = T_k(v) T_k(u).
\end{split}
\end{align}
    For $\CalQ_N[F]$, we simply take the limit $m_k \to \infty$ if $k \notin F$. 
\hfill $\square$

\subsection{Proof of \cref{lem:QQ}} \label{proof:lem:QQ}

The lemma is proved by induction. 
The commutativity is immediate for $k=1$. 
By the pentagon identity \eqref{eq:qdf-pentagon}, we have
\begin{align*}
    T_{k+1}(u)T_k(v) = T_k(v) U_k(u+v) T_{k+1}(u),
\end{align*}
where we define an operator
\begin{align*}
\begin{split}
    U_k(u) & = \varphi(p_k+x_{k+1}-x_{k-1}+u-m_{k+1})\varphi(p_k+x_{k+1}-x_k +u- m_{k+1}) \\
    & \quad \times \varphi(p_k+p_{k+1}+x_{k+1}-x_{k-1}+u)\varphi(p_k+p_{k+1}+x_{k+1}-x_k+u).
\end{split}
\end{align*}
Then, assuming that it holds for $k$, we have
\begin{align}\nonumber
\begin{split}
    \bQ_{k+1}(u) \bQ_{k+1}(v) & = \bQ_{k}(u) T_{k+1}(u) \bQ_{k-1}(v) T_k(v) T_{k+1}(v) \\
    & = \bQ_k(u) \bQ_{k-1}(v) T_{k+1}(u)T_k(v)T_{k+1}(v) \\
    & = \bQ_{k}(u) \bQ_{k-1}(v) T_k(v) U_k(u+v) T_{k+1}(u) T_{k+1}(v) \\
    & = \bQ_{k}(v) \bQ_{k-1}(u) T_k(u) U_k(u+v) T_{k+1}(v) T_{k+1}(u) \\
    & = \bQ_k(v) \bQ_{k-1}(u) T_{k+1}(v) T_k(u) T_{k+1}(u) \\
    & = \bQ_k(v) T_{k+1}(v) \bQ_{k-1}(u) T_k(u) T_{k+1}(u) \\
    & = \bQ_{k+1}(v) \bQ_{k+1}(u) ,
\end{split}
\end{align}
from which we conclude the proof.
\hfill $\square$

%\vspace{1cm}
%%%%%% Bibliography %%%%%%
\bibliography{Whittaker} 
\bibliographystyle{ytamsalpha}

\end{document}